\documentclass[english,aps,pra,two column,floatfix]{revtex4-1}
\usepackage[T1]{fontenc}
\usepackage[latin9]{inputenc}
\usepackage{color}
\usepackage{bm}
\usepackage{amsmath}
\usepackage{amssymb}
\usepackage{graphicx}
\usepackage{esint}

\makeatletter

\@ifundefined{textcolor}{}
{%
 \definecolor{BLACK}{gray}{0}
 \definecolor{WHITE}{gray}{1}
 \definecolor{RED}{rgb}{1,0,0}
 \definecolor{GREEN}{rgb}{0,1,0}
 \definecolor{BLUE}{rgb}{0,0,1}
 \definecolor{CYAN}{cmyk}{1,0,0,0}
 \definecolor{MAGENTA}{cmyk}{0,1,0,0}
 \definecolor{YELLOW}{cmyk}{0,0,1,0}
}

\makeatother

\usepackage{babel}
\begin{document}

\title{Quantum probabilistic sampling of multipartite 60-qubit Bell inequality
violations}

\author{M. D. Reid$^{1}$, B. Opanchuk$^{1}$, L. Rosales-Z\'arate$^{1}$,
P. D. Drummond$^{1}$}

\affiliation{$^{1}$Centre for Quantum and Optical Science, Swinburne University
of Technology, Melbourne 3122, Australia}
\begin{abstract}
We show that violation of genuine multipartite Bell inequalities can
be obtained with sampled, probabilistic phase space methods. These
genuine Bell violations cannot be replicated if any part of the system
is described by a local hidden variable theory. The Bell violations
are simulated probabilistically using quantum phase-space representations.
We treat mesoscopically large Greenberger-Horne-Zeilinger (GHZ) states
having up to $60$ qubits, using both a multipartite SU(2) Q-representation
and the positive P-representation. Surprisingly, we find that sampling
with phase-space distributions can be exponentially\emph{ faster}
than experiment. This is due to the classical parallelism inherent
in the simulation of quantum measurements using phase-space methods.
Our probabilistic sampling method predicts a contradiction with local
realism of ``Schr\"odinger-cat'' states that can be realized as
a GHZ spin state, either in ion traps or with photonic qubits. We
also present a quantum simulation of the observed super-decoherence
of the ion-trap ``cat'' state, using a phenomenological noise model.
\end{abstract}


\maketitle

\section{Introduction}

Quantum simulation of systems with many degrees of freedom is a difficult
and interesting problem of much topical interest. Calculating the
dynamics of many-body quantum systems is hard, since the Hilbert space
dimension increases exponentially with the number of modes or degrees
of freedom~\cite{Dirac1929,Feynman1982}. There are two main approaches:
one can do a computational simulation~\cite{Haake1979,Corney2006,Deuar2007,Alon2008,Gambetta2008,Trotzky2012},
or else a physical simulation with another quantum system~\cite{Cirac2003,Jaksch2005,Buluta2009,Islam2011,Georgescu2014}.
Universal quantum computers provide a third option~\cite{Lanyon2011},
but these are limited in size.

One path to solving this problem is to use probabilistic simulations
whose correlations correspond to quantum averages. For large problems,
this approach was pioneered by Glauber and co-authors~\cite{Glauber1978,Haake1979},
who studied quantum statistics of super-fluorescence. Later, their
approximate method was generalized to an exact probabilistic representation
of arbitrary quantum states~\cite{Drummond1980,Carter1987}. Quantum
simulation predictions were experimentally verified for multi-mode
optical fields displaying squeezing and quantum entanglement~\cite{Rosenbluh1991,Drummond1993-solitons,Heersink2005,Corney2006}.
More recently, the method has been applied to colliding BEC systems~\cite{Deuar2007,Lewis-Swan2014},
and to Bell violations in parametric down-conversion experiments~\cite{Rosales-Zarate2014}. 

Here we study how efficiently such probabilistic methods can be used
to simulate the most extreme quantum superposition states $-$ or
``Schr\"odinger cat'' states. The ``cat'' state is often represented
as a GHZ state for $M$ particles~\cite{Greenberger1989}: 
\begin{equation}
\vert\Phi\rangle=\frac{1}{\sqrt{2}}\left(\vert\uparrow\ldots\uparrow\rangle+e^{i\phi}\vert\downarrow\ldots\downarrow\rangle\right).\label{eq:ghz-full-1}
\end{equation}
where $|\uparrow\ldots\uparrow\rangle=\bigotimes_{j=1}^{M}|\uparrow\rangle_{j}$,
$|\downarrow\ldots\downarrow\rangle=\bigotimes_{j=1}^{M}|\downarrow\rangle_{j}$
and $|\uparrow\rangle_{j}$, $|\downarrow\rangle_{j}$ are the eigenstates
of the spin $\hat{\sigma}^{z}$ of the $j$-th particle. A powerful
signature of the ``cat'' state is its $M$-qubit nonlocality. These
have been explored in photonic~\cite{Lu2007} and ion trap experiments~\cite{Leibfried2005-creation},
which demonstrated Bell-Mermin violations and genuine $M$-particle
entanglement for up to $M=14$\textcolor{red}{{} }ions~\cite{Lanyon2014}. 

We investigate probabilistic methods for simulating both Bell-Mermin
violations and the more challenging Svetlichny-Collins genuine Bell
violations in these multipartite ``cat'' states. The latter inequality
allows us to demonstrate the genuine $M$-partite nonlocality of the
multipartite GHZ state (\ref{eq:ghz-full-1}) for up to $M=60$ ions
or modes. This is a true Schr\"odinger cat signature: it cannot be
obtained if any subset has a local hidden variable (LHV) description.
It is often thought that probabilistic sampling would be extraordinarily
difficult for a ``cat'' state of large size. Probabilistic methods
using measured eigenvalues are impossible, since this would amount
to an LHV theory, which cannot violate a Bell inequality. 

Importantly, Mermin showed that the difference between LHV predictions
and quantum predictions scales exponentially with increasing system
size $M$~\cite{Mermin1990-entanglement}, making this a significant
challenge for probabilistic methods. However, the techniques used
in this paper do \emph{not} rely on LHV theories, but instead sample
over stochastic variables whose values are permitted to go beyond
the eigenvalue spectrum. The ability to simulate quantum mechanics
in this way gives a beautiful analogy to the theory of weak values
and measurements~\cite{Aharonov1988}, as has been explained elsewhere~\cite{Drummond2014-bell-sim,Opanchuk2014-bell-sim,Rosales-Zarate2014}. 

We find that sampling errors of high-order correlations are larger
than for low-order correlations. We note that $M$-order correlations
are needed to display the signature of an $M$-partite Bell-Svetlichny
nonlocality. However, for these results, phase-space simulations have
a classical parallelism not available in any quantum experiment. This
parallelism occurs because exponentially many non-commuting measurements
can be calculated at once. The result is an exponential \emph{speedup}
for simulating multipartite Bell violations. The simulation is $e^{M/3}$
times faster than experiment with methods used here.

The general advantage of probabilistic sampling in quantum simulations
compared to wave-function methods~\cite{Islam2011} is that the required
computational memory scales linearly, not exponentially, with the
number of qubits. This eliminates the problem of exponential scaling
in memory size found in direct, orthogonal basis calculations. The
time taken, or equivalently the number of samples required, depends
on the type of measurement and the resulting sampling error in a finite
ensemble. We consider the Greenberger-Horne-Zeilinger (GHZ) state
(\ref{eq:ghz-full-1}), and simulate spin correlations as well as
multipartite Bell violations.

The utility of phase-space representations is that they provide a
route to performing such probabilistic sampling. We employ two common
positive phase-space distributions, namely, the SU(2) Q function~\cite{Husimi1940,Arecchi1972,Gilmore1975}
and the positive P-distribution~\cite{Drummond1980}. The latter
method has already been used to obtain analytic results for probabilistic
Bell violation~\cite{Drummond1983}. To focus on the sampling issue,
we mostly treat static cases with known probability distributions.
We also treat dynamical simulations of decoherence. A summary of the
results is published elsewhere~\cite{Opanchuk2014-bell-sim}.

In these multipartite investigations, we use Bell-like inequalities
that extend the usual bipartite inequalities to many qubits. We test
the MABK (Mermin-Ardehaly-Belinski-Klyshko) inequality~\cite{Mermin1990-entanglement,Ardehali1992,Belinskii1993-interference,Belinsky1993-N-particle}
which is a Bell inequality generalized to multipartite qubit systems,
and the Collins-Svetlichny inequalities~\cite{Svetlichny1987,Collins2002},
which are sufficient conditions for genuine multipartite Bell violations~\cite{Ghose2009,Ajoy2010,Bancal2011,Chen2011,Grandjean2012}.
Genuine multipartite Bell violations prove that Bell violations are
a macroscopic property.

We find different behavior depending on the order of the correlation
function. There is no growth in sampling error with the number of
qubits when simulating low-order correlations in our calculations.
Thus, fixed order correlations do not have an exponential increase
in simulation time. However, correlations with a growing order equal
to the number of qubits take an exponentially long time to simulate.
Yet even these calculations scale only as a \emph{fractional} power
of the number of qubits. This allows us to simulate genuine multipartite
Bell violations of GHZ states with 60 qubits, corresponding to a Hilbert
space of a quintillion ($10^{18}$) dimensions. 

Bell violations as large as this would also require a quintillion
different measurement settings in the laboratory. At around 1 ms per
measurement in an ion-trap experiment, full confirmation of a 60 qubit
multipartite Bell violation would take over $30$ million years, even
with just one measurement per laboratory setting. For multipartite
Bell inequalities, our simulations took less than $48$ hours, so
the exponential speedup obtained through phase space quantum simulations
is a highly practical computational tool. To demonstrate applications
for decoherence dynamics, we use the method to simulate the observed
super-decoherence in ion traps.

The paper is organized as follows. In Section~\ref{sec:MABK and N Partite Bell Ineq}
we discuss the multipartite Bell inequalities. The sampling of the
GHZ states using the positive P and the Q section are described in
Sections~\ref{sec:Sampling-GHZ-states} and~\ref{sec:Sampling-GHZ-Qfunction},
respectively. The computational results are shown in Section~\ref{sec:Results}.
In Section~\ref{sec:Decoherence-model} we describe a decoherence
model that shows the dynamical decay of the Bell inequality, as observed
experimentally. Finally, Section~\ref{sec:Conclusions} gives a summary
of our results and conclusions.

\section{\label{sec:MABK and N Partite Bell Ineq} Multipartite Bell inequalities }

The challenge for quantum simulation is to simulate very large systems
where quantum effects can still manifest themselves. The best example
is a macroscopic superposition state of the type considered in the
``Schrödinger cat'' paradox. With this objective, we will analyze
how to simulate the genuine multipartite Bell inequality violations
of $M$ entangled particles. Our goal is to determine whether this
is possible, using probabilistic sampling. We also wish to understand
the relevant scaling properties, as they depend on the measurements
themselves. A detailed treatment of the bipartite case, including
dynamical simulations, is presented elsewhere~\cite{Rosales-Zarate2014}.

\subsection{MABK Bell inequalities for $M$ sites}

First, we summarize well-known Bell inequalities that test local hidden
variable (LHV) theories involving $M$ spin-$1/2$ particles at different
sites. We label the sites by $j$, where $j=1,\ldots,M$. 

In the case of $M$ particles emitted from a common source, measurements
of $M$ spatially separated observers are modeled in the LHV theory
by taking random samples of a common set of parameters (the hidden
variables) symbolized by $\lambda$. Measured values are then functions
of some local detector/analyzer settings and the hidden parameters
$\lambda$. 

We use the notation that $X_{m}(\lambda)\equiv X_{m}(\theta_{m},\lambda)$
for the $m$-th observer with the detector analyzer setting $\theta_{m}$,
denoting the measurement value by $X_{m}$. Here, the measurement
event includes the selection of the measurement setting $\theta_{m}$
at each site. The $M$ measurement events are assumed to be space-like
separated. In an LHV theory the correlations are thus obtained from
a probabilistic calculation of the form:
\begin{eqnarray}
E(X_{1},X_{2},\ldots X_{M}) & \equiv & \left\langle \left[\prod_{m=1}^{M}X_{m}\right]\right\rangle \nonumber \\
 & = & \int\left[\prod_{m=1}^{M}X_{m}(\lambda)\right]P(\lambda)d\lambda.\label{eq:LHV}
\end{eqnarray}
where $P(\lambda)$ is a probability distribution for the hidden variables
$\lambda$. 

One can consider that at each site the experimentalist makes one of
two choices for the measurement. Here, we denote these two choices
by the quantum observables $\hat{x}_{j}$ and $\hat{y}_{j}$, and
denote the outcomes associated with these measurements by $X_{j}$,
$Y_{j}$ respectively. Experimentally, one uses an adjustable polarizer
or Rabi rotation at each site to determine which of the choices to
make, and there are $2^{M}$ possible combinations. For each of these
$2^{M}$ choices, an ensemble of measurements is necessary to obtain
the relevant correlations.

Following Mermin~\cite{Mermin1990-entanglement}, we can construct
for mathematical convenience the operator
\begin{equation}
\hat{A}_{j}=\hat{x}_{j}+i\hat{y}_{j},\label{eq:f}
\end{equation}
bearing in mind that this is not a measured observable. We can also
define the complex function $F_{j}=X_{j}+iY_{j}$. We now examine
the dichotomic case using qubits. We follow Mermin and choose:
\begin{eqnarray}
\hat{x}_{j} & = & \hat{\sigma}_{j}^{\theta_{j}}\nonumber \\
\hat{y}_{j} & = & \hat{\sigma}_{j}^{\theta_{j}+\pi/2},
\end{eqnarray}
 where $\hat{\sigma}_{j}^{\theta}=\hat{\sigma}_{j}^{x}\cos\theta_{j}+\hat{\sigma}_{j}^{y}\sin\theta_{j}$,
and $\hat{\sigma}_{j}^{x/y}$ are the Pauli spin operators. Therefore:
\begin{equation}
\hat{A}_{j}=\left(\hat{\sigma}_{j}^{x}+i\hat{\sigma}_{j}^{y}\right)e^{-i\theta_{j}}.
\end{equation}

Next, we consider the measurable moments given by the expression:
\begin{equation}
A_{\mathrm{QM}}=\langle\prod_{j=1}^{M}\hat{A}_{j}\rangle\equiv\left\langle \hat{A}\right\rangle ,\qquad\hat{A}\equiv\prod_{j=1}^{M}\hat{A}_{j}
\end{equation}
and the corresponding LHV prediction for this moment

\[
A_{\lambda}=\langle\prod_{j=1}^{M}\hat{A}_{j}\rangle_{\lambda}\equiv\langle\prod_{j=1}^{M}F_{j}\rangle.
\]
where $\Pi$ denotes the product (standard notation). One can expand
the terms of the product, and write as a real and imaginary part:
So, we define the real and imaginary parts by:
\begin{equation}
A_{\mathrm{QM}/\lambda}=\mathrm{Re}A_{\mathrm{QM}/\lambda}+i\mathrm{Im}A_{\mathrm{QM}/\lambda}.\label{eq:expfreim}
\end{equation}

It is known that LHV theories place a constraint on what should be
observed for these quantities. These are the Mermin-Ardehali-Belinski\u{\i}-Klyshko
(MABK) Bell inequalities. Mermin~\cite{Mermin1990-entanglement}
originally derived the following Bell inequality (which we will call
Mermin's inequality),

\begin{equation}
\mathrm{Im}A_{\lambda}\le\begin{cases}
2^{(M-1)/2}, & M\ \mathrm{is\ odd},\\
2^{M/2}, & M\ \mathrm{is\ even}.
\end{cases}\label{eq:MABKMermin}
\end{equation}
The same inequalities hold for the $\mathrm{Re}A_{\lambda}$. Mermin's
inequality for even $M$ is weak, and is not violated by the Bell
state~(\ref{eq:ghz}) for $M=2$. Therefore for the case of even
$M$ we will follow Ardehali, Belinski\u{\i} and Klyshko (ABK)~\cite{Ardehali1992,Belinskii1993-interference,Belinsky1993-N-particle},
who derived the following inequalities: 

\begin{equation}
\mathrm{Re}A_{\lambda}+\mathrm{Im}A_{\lambda}\le\begin{cases}
2^{M/2}, & M\ \mathrm{is\ even},\\
2^{(M+1)/2}, & M\ \mathrm{is\ odd}.
\end{cases}\label{eq:MABKArdehali}
\end{equation}
ABK inequalities are stronger for even $M$, but not for odd $M$,
and thus the MABK Bell inequalities~\cite{Belinskii1993-interference,Belinsky1993-N-particle}
are the combination of~(\ref{eq:MABKMermin}) for odd $M$, and~(\ref{eq:MABKArdehali})
for even $M$. 

We can expand these inequalities explicitly to see what they are.
For $M=2$, $\theta_{j}=0$, the MABK inequality is:
\begin{equation}
\langle\sigma_{1}^{x}\sigma_{2}^{y}\rangle_{\lambda}+\langle\sigma_{1}^{y}\sigma_{2}^{x}\rangle_{\lambda}+\langle\sigma_{1}^{x}\sigma_{2}^{x}\rangle_{\lambda}-\langle\sigma_{1}^{y}\sigma_{2}^{y}\rangle_{\lambda}\leq\sqrt{2}\label{eq:sigma2}
\end{equation}
which is the famous Clauser-Horne-Shimony-Holt (CHSH) Bell inequality.
For $M=3$, $\theta_{j}=0$ the resulting inequality is:
\begin{equation}
\langle\sigma_{1}^{y}\sigma_{2}^{x}\sigma_{3}^{x}\rangle_{\lambda}+\langle\sigma_{1}^{x}\sigma_{2}^{y}\sigma_{3}^{x}\rangle_{\lambda}+\langle\sigma_{1}^{x}\sigma_{2}^{x}\sigma_{3}^{y}\rangle_{\lambda}-\langle\sigma_{1}^{y}\sigma_{2}^{y}\sigma_{3}^{y}\rangle_{\lambda}\leq2\label{eq:sigma2-1}
\end{equation}
as derived by Mermin. We note by defining $F_{j}=X_{j}-iY_{j}$ a
different set of MABK inequalities with different signs can be derived.

\subsection{MABK violations with a GHZ state}

All of the MABK inequalities are predicted by LHV theories, but only
for the right quantum state are they maximally violated. Let us consider
the GHZ state:
\begin{equation}
|\psi\rangle=\frac{1}{\sqrt{2}}\left(\bigotimes_{j=1}^{M}|\uparrow\rangle_{j}+e^{i\phi}\bigotimes_{j=1}^{M}|\downarrow\rangle_{j}\right),\label{eq:ghz}
\end{equation}
where $|\uparrow\rangle_{j}$ $|\downarrow\rangle_{j}$ are the eigenstates
of $\hat{\sigma}_{j}^{z}$. \textcolor{red}{} It is known that the
state~(\ref{eq:ghz}) with $r=M$ violates~(\ref{eq:MABKArdehali})
by the \emph{maximum} amount predictable by Quantum Mechanics (QM)~\cite{Werner2001}.
For the Mermin-type inequalities~(\ref{eq:MABKMermin}), this maximal
violation occurs for the angle $\phi=\pi/2$ and the measurement choice
$\theta_{j}=0$:
\begin{equation}
A_{j}=\sigma_{j}^{x}+i\sigma_{j}^{y},\quad j=1,\ldots,M.\label{eq:angleperfect}
\end{equation}
 where we have now denoted the results $X_{j}$, $Y_{j}$ of the measurements
$\hat{\sigma}_{j}^{x}$, $\hat{\sigma}_{j}^{y}$ by $\sigma_{j}^{x}$,
$\sigma_{j}^{y}$ written without the operators. This orthogonal angle
choice corresponds to the famous cases of the EPR-Bohm and GHZ paradoxes~\cite{Einstein1935,Greenberger1989,Mermin1990-reality},
that yield perfect correlations between spatially separated spins.
The quantum prediction for the choice of measurement orientations~(\ref{eq:angleperfect})
is~\cite{Mermin1990-entanglement}:
\begin{equation}
\mathrm{Im}A_{\mathrm{QM}}=2^{M-1}.\label{eq:QMMermin}
\end{equation}
On the other hand, the Ardehali-Bell-CHSH-type inequalities~(\ref{eq:MABKArdehali})
give a maximum when $\phi=\pi$ and one site has a shifted measurement
angle: 
\begin{eqnarray}
F_{j} & = & \sigma_{j}^{x}-i\sigma_{j}^{y},\,\, j\neq M\label{eq:anglenonper}\\
F_{M} & = & \sigma^{-\pi/4}+i\sigma^{\pi/4}.\nonumber 
\end{eqnarray}
We note this corresponds for $M=2$ to the case of Bell and CHSH~\cite{Bell1964,Clauser1969,Clauser1978,D'Espagnat1971}.
Here, the measurement choice does not allow perfect correlation between
spatially separated measurements for a fixed setting, and the violation
is obtained statistically. The quantum prediction in this case is~\cite{Ardehali1992,Mermin1990-entanglement}:
\begin{equation}
\mathrm{Re}A_{\mathrm{QM}}+\mathrm{Im}A_{\mathrm{QM}}=2^{M-1/2}.\label{eq:QMArdehali}
\end{equation}

It is convenient to join the odd- and even-$M$ inequalities using
an operator

\begin{equation}
\hat{V}=\begin{cases}
\mathrm{Re}\hat{A}+\mathrm{Im}\hat{A}, & M\ \mathrm{is\ even},\\
\sqrt{2}\,\mathrm{Im}\hat{A}, & M\ \mathrm{is\ odd}.
\end{cases}\label{eq:V_Op}
\end{equation}
In this case the MABK inequality for all $M$ is, in the case of an
LHV theory:

\begin{equation}
V_{\lambda}\equiv\vert\langle\hat{V}\rangle_{\lambda}\vert\le2^{M/2}.\label{eq:V}
\end{equation}
This is violated by quantum mechanics with the state and measurement
choices above, since:

\begin{equation}
V_{\mathrm{QM}}\equiv\langle\hat{V}\rangle=2^{M-1/2}>V_{\lambda}.
\end{equation}
The ratio between the LHV limit and the QM result is thus:

\begin{equation}
\frac{V_{\mathrm{QM}}}{V_{\lambda}}\ge2^{\left(M-1\right)/2},
\end{equation}
which grows exponentially with $M$.

\subsection{Genuine $M$-partite Bell nonlocality}

Svetlichny~\cite{Svetlichny1987} introduced the idea of genuine
multipartite nonlocality. He derived inequalities that if violated
indicate a three-body (rather than two-body) nonlocality. The inequalities
have been generalized to $M-$partite cases by Collins \textit{et
al}~\cite{Collins2002} and by Seevinck and Svetlichny~\cite{Seevinck2002}.
We point out that other recent works~\cite{Aolita2012,Gallego2012,Bancal2013}
have improved Svetlichny's approach further. 

The Svetlichny-CGPRS inequality is:
\begin{eqnarray}
V_{{\cal S}}\equiv\mathrm{Re}A_{\lambda}+\mathrm{Im}A_{\lambda} & \leq & 2^{M-1}\,,\label{eq:merminsteerstat-1}
\end{eqnarray}
the violation of which is sufficient to confirm \emph{genuine $M$-partite
Bell nonlocality}. For $M=3$ this means that the violation cannot
be explained using product states or mixtures with Bell nonlocality
between only two sites. More generally, for arbitrary $M$, this terminology
means that the violation cannot be explained using states with a genuine
$ $$m$-partite Bell nonlocality, where $m<M$. The quantum prediction
maximizes at~(\ref{eq:QMArdehali}) to predict violation, for even
$M$, by a \emph{constant }amount: 
\begin{equation}
\frac{V_{QM}}{V_{S}}=\sqrt{2}\,.
\end{equation}

This constant violation ratio differs from the exponential violation
predicted for the MABK inequalities, which makes the effect both harder
to measure experimentally, and more difficult to simulate than the
usual Bell inequality. However it is necessary to achieve this stronger
correlation if one wishes to assert that a given superposition is
truly macroscopic to a given level, i.e., if one wishes to exclude
the possibility that there are only microscopic violations of local
realism present in a quantum system.

\textcolor{red}{}

\section{\label{sec:Sampling-GHZ-states}Sampling GHZ states with positive
phase-space distributions}

The states we wish to sample are GHZ states~(\ref{eq:ghz}), which
are experimentally prepared in a number of photonic and ion-trap experiments.
We rewrite these as:
\begin{equation}
\vert\Phi\rangle=\frac{1}{\sqrt{2}}\left(\vert\uparrow\ldots\uparrow\rangle+e^{i\phi}\vert\downarrow\ldots\downarrow\rangle\right).\label{eq:ghz-full}
\end{equation}
Of course, any experiment will inevitably also include other states
owing to decoherence effects. Here we wish to start by considering
the pure state, which is a worst-case scenario from the point of view
of phase-space simulations. The up- and down-states can be represented
differently, depending on the underlying physical system, which will
in turn affect the sampling. We will consider different sampling techniques
using different operator representations, in order to compare their
efficiency.

\subsection{Phase-space methods}

In general terms, a phase-space representation is a mapping from a
c-number distribution function $P\left(\vec{\lambda}\right)$ to a
density matrix $\hat{\rho}$, defined by

\begin{equation}
\hat{\rho}=\int P\left(\vec{\lambda}\right)\hat{\Lambda}\left(\vec{\lambda}\right)d\vec{\lambda}.\label{eq:general-phase-space}
\end{equation}

Here $\hat{\Lambda}\left(\vec{\lambda}\right)$ is a complete operator
basis, which is parametrized with a phase-space variable $\vec{\lambda}$,
and $P\left(\vec{\lambda}\right)$ is a distribution over $\vec{\lambda}$
which typically allows one to calculate observables as moments. For
our present purposes, we will focus on mappings that involve a positive-definite
distribution $P\left(\vec{\lambda}\right)$. This allows probabilistic
sampling, which is a very scalable route for calculating high-dimensional
integrals and correlations. It also removes the need to have a numerical
representation of an exponentially large matrix. This approach results
in efficient scaling for low-order correlations, even for highly nonclassical
states like the GHZ state, and can be sampled for high-order correlations
with somewhat lower efficiency. 

There are many such mappings known. The earliest methods developed
were for the Wigner function~\cite{Wigner1932}, Q-function~\cite{Husimi1940}
and P-function~\cite{Glauber1963-states,Sudarshan1963}. These are
all for bosonic Hilbert spaces, are defined for a real phase-space
$\vec{\lambda}$, and correspond to different operator orderings.
Of these, only the Q-function is positive-definite. Subsequently,
positive-definite extensions of these were developed that use complex
instead of real phase-spaces, including the positive P-representation~\cite{Drummond1980},
the positive Wigner representation~\cite{Chaturvedi1994,DeOliveira1992}
and the Gaussian representation~\cite{Corney2003}. The positive
P-representation is useful, as it combines stochastic time-evolution
with simple observables.

All these bosonic methods involve a Hilbert space of too large a dimension
for optimum sampling of the GHZ state, as we explain below. It is
most efficient to only represent those parts of a Hilbert space that
are measured. Hence, it is better to use a phase-space representations
that is specifically matched to a finite dimensional Hilbert space.
The earliest of these were the SU(2) based continuous representations~\cite{Radcliffe1971,Arecchi1972,Zhang1990},
which employ Lie group methods. These have a similar form to the bosonic
case. Once again, there are both positive and non-positive distributions,
as well as complex phase-space methods~\cite{Barry2008}. A widely
used positive form is the SU(2) Q-function~\cite{Arecchi1972,Gilmore1975},
which we analyze in detail in the next section.

Recently, a number of interesting and innovative methods have been
introduced that treat finite Hilbert spaces in a different way. These
replace the integral in Eq (\ref{eq:general-phase-space}) with a
summation over a finite set of points. Using this technique, it is
possible to develop a discrete Wigner distribution~\cite{Wootters1987,Wootters1989,Gibbons2004,Wootters2004,Wootters2006,Bjork2008},
which uses hermitian matrices instead of distributions to represent
the Hilbert space. In the standard construction of such methods, certain
specific quantum states have positive representations, but this is
not true in general. In other words, the generic case for the discrete
Wigner distribution is that the mapping is non-positive.

These discrete approaches have the property that the underlying discrete
Wigner distribution is a $2^{M}\times2^{M}$ matrix for $M$ qubits
\cite{Wootters2004,Gibbons2004,Bjork2008}. In the largest case treated
here, with $M=60$, this involves $10^{36}$ matrix elements. These
do not all have to be stored in memory, which is impossible with current
computers. Nevertheless, calculating observables with $10^{36}$ elements
requires sampling to reduce the computation time. As the elements
are not all positive, this would presumably involve a sign or phase
term, which can lead to inefficiencies. 

Accordingly, we do not investigate the discrete Wigner function here.
Yet such methods may also be useful. The main challenge is that the
resulting large matrix representations are not probabilistic. The
question of how to sample these efficiently is an open question at
present. However, extending such discrete techniques to allow probabilistic
sampling may not be impossible. This is outside the scope of the present
paper, so we now return to the question of efficient sampling using
continuous, positive phase-space distributions.

\subsection{Positive P-representation}

We first consider the positive P-representation~\cite{Drummond1980}.
This is a probabilistic phase-space representation widely used in
quantum optics. It is most suitable when using photonic methods to
obtain qubit observables, as it can represent any multi-mode bosonic
quantum state. With this representation, a general quantum density
matrix $\hat{\rho}$ is represented using a positive distribution
$P\left(\vec{\alpha},\vec{\beta}\right)$, where:

\begin{equation}
\hat{\rho}=\int P\left(\vec{\alpha},\vec{\beta}\right)\hat{\Lambda}\left(\vec{\alpha},\vec{\beta}\right)d^{2M}\vec{\alpha}d^{2M}\vec{\beta}.
\end{equation}
Here the projector $\hat{\Lambda}$ is:
\begin{equation}
\hat{\Lambda}\left(\vec{\alpha},\vec{\beta}\right)=\frac{\left|\vec{\alpha}\right\rangle \langle\vec{\beta}^{*}\vert}{\langle\vec{\beta}^{*}\vert\vec{\alpha}\rangle},
\end{equation}
where $\left|\vec{\alpha}\right\rangle =\left|\alpha_{1}.\ldots\alpha_{n}\right\rangle $
is a multi-mode coherent state. 

This representation maps quantum states into $4M$ real coordinates:\textbf{
$\vec{\alpha},\vec{\beta}$}, which is twice the dimension of a classical
phase-space. The expectation of any normally ordered observable $\hat{O}\equiv O(\hat{a}_{1}^{\dagger},\hat{a}_{1},\ldots)$
is then:
\begin{equation}
\left\langle \hat{O}\right\rangle =\int O(\beta_{1},\alpha_{1},\ldots)P(\vec{\alpha},\vec{\beta})d^{2M}\vec{\alpha}\, d^{2M}\vec{\beta}.\label{+Pcorrels}
\end{equation}

A general, although non-unique positive construction is:
\begin{equation}
\begin{split}P(\vec{\alpha},\vec{\beta})= & \frac{\left\langle \vec{\mu}\right|\widehat{\rho}\left|\vec{\mu}\right\rangle }{\left(2\pi\right)^{2M}}e^{-\left|\vec{\nu}\right|^{2}}\end{split}
\,,\label{eq:P-from-rho}
\end{equation}
where we have made a variable change to sum and difference variables:
\begin{equation}
\vec{\nu}=\left(\vec{\alpha}-\vec{\beta}^{*}\right)/2,\quad\vec{\mu}=\left(\vec{\alpha}+\vec{\beta}^{*}\right)/2\,.\label{eq:P-variable-change}
\end{equation}

\subsection{Spin state representation}

The natural choice for up- and down-states are spin states $\vert\uparrow\rangle\equiv\vert10\rangle$,
$\vert\downarrow\rangle\equiv\vert01\rangle$. Spin operators can
be mapped into bosons with the Schwinger representation~\cite{Biedenharn1965}:
\begin{eqnarray}
\sigma_{j}^{x} & = & \hat{a}_{j}^{\prime\dagger}\hat{a}_{j}^{\prime\prime}+\hat{a}_{j}^{\prime\prime\dagger}\hat{a}_{j}^{\prime},\nonumber \\
\sigma_{j}^{y} & = & \frac{1}{i}\left(\hat{a}_{j}^{\prime\dagger}\hat{a}_{j}^{\prime\prime}-\hat{a}_{j}^{\prime\prime\dagger}\hat{a}_{j}^{\prime}\right),\nonumber \\
\sigma_{j}^{z} & = & \hat{a}_{j}^{\prime\dagger}\hat{a}_{j}^{\prime}-\hat{a}_{j}^{\prime\prime\dagger}\hat{a}_{j}^{\prime\prime},
\end{eqnarray}
where $\hat{a}_{j}^{\prime\dagger}$ creates a particle in the first
position of the $j$-th spin operator, and $\hat{a}_{j}^{\prime\prime\dagger}$
creates one in the second position. Substituting $\hat{\rho}=\vert\Phi\rangle\langle\Phi\vert$
into~(\ref{eq:P-from-rho}) and performing the substitution~(\ref{eq:P-variable-change}),
we get the following positive-P function:
\begin{equation}
P=\frac{1}{2\pi^{4M}}e^{-\vert\vec{\nu}\vert^{2}}e^{-\vert\vec{\mu}^{\prime}\vert^{2}-\vert\vec{\mu}^{\prime\prime}\vert^{2}}\left|\prod_{j=1}^{M}\mu_{j}^{\prime}+e^{-i\phi}\prod_{j=1}^{M}\mu_{j}^{\prime\prime}\right|^{2}.
\end{equation}

To sample this distribution, we use the von Neumann rejection method,
which requires a known reference distribution as an upper bound. This
distribution is bounded above by the following expression:
\begin{equation}
P\le2G(\vec{\nu})P_{0}(\vec{\mu})\,,
\end{equation}
where:
\begin{equation}
G(\vec{\nu})=\frac{1}{\pi^{2M}}e^{-\vert\vec{\nu}\vert^{2}},
\end{equation}
and
\begin{equation}
P_{0}=\frac{1}{2\pi^{2M}}e^{-\vert\vec{\mu}^{\prime}\vert^{2}-\vert\vec{\mu}^{\prime\prime}\vert^{2}}\left(\prod_{j=1}^{M}\left|\mu_{j}^{\prime}\right|^{2}+\prod_{j=1}^{M}\left|\mu_{j}^{\prime\prime}\right|^{2}\right).
\end{equation}

These two reference distributions can be sampled exactly using a combination
of Gamma and Gaussian variates. The expectation of the Mermin operator
$\hat{A}$ of interest here is then given by:
\begin{eqnarray}
\langle\hat{A}\rangle & = & \int d^{4M}\vec{\alpha}d^{4M}\vec{\beta}P\left(\vec{\mu}(\vec{\alpha},\vec{\beta}),\vec{\nu}(\vec{\alpha},\vec{\beta})\right)\nonumber \\
 &  & \times\prod_{j=1}^{M}\left(\left(\beta_{j}^{\prime}\alpha_{j}^{\prime\prime}+is_{j}\beta_{j}^{\prime\prime}\alpha_{j}^{\prime}\right)e^{-is_{j}\theta_{j}}\right).
\end{eqnarray}

While this method is able to sample the required GHZ state, the sampling
is rather inefficient. We can improve the results using a more compact
Hilbert space mapping technique, described in the next subsection.

\subsection{Number state representation}

Sampling is generally improved if the Hilbert space dimension is reduced
as far as possible, to eliminate samples that overlap the unused part
of the space. We can decrease the number of dimensions in the required
phase space by half, by using number states instead of spin states.
This is possible because we really only need the fact that occupations
are binary.

\begin{figure}
\includegraphics{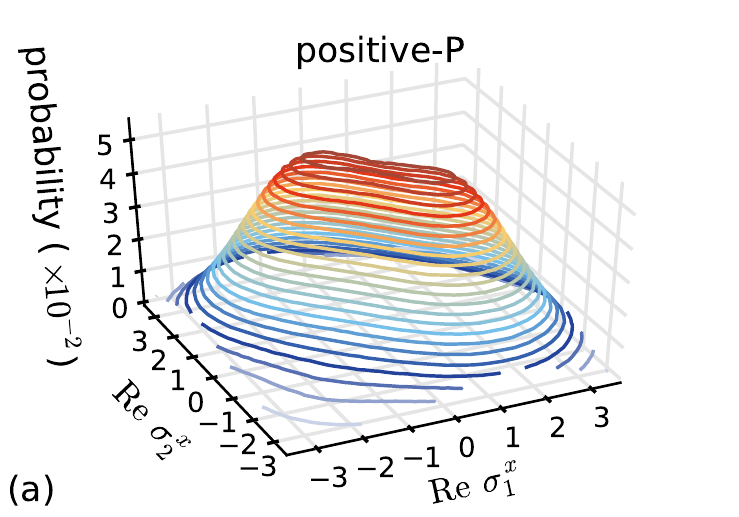}\\
\includegraphics{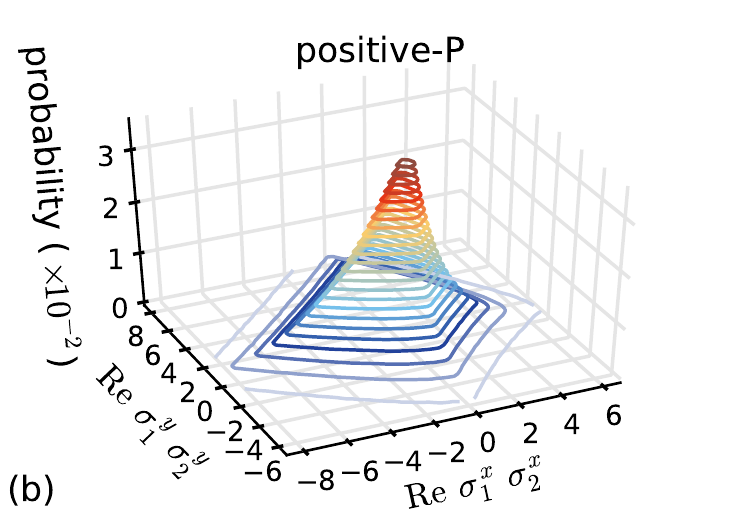}

\protect\caption{Correlations for the different parts of the quantity~(\ref{eq:F-2-particles-1})
in the positive-P representation, with the number state method and
$2^{26}$ samples.\label{fig:Spin-moments-P}}
\end{figure}

This can be done because our operators of interest \textemdash{} $\hat{A}$
\textemdash{} depend on $\sigma_{j}^{x}$ and $\sigma_{j}^{y}$ linearly.
Therefore if we denote $\vert\uparrow\rangle\equiv\vert1\rangle$,
$\vert\downarrow\rangle\equiv\vert0\rangle$, we can formally write:
\begin{eqnarray}
\sigma_{x}^{j} & = & \hat{a}_{j}+\hat{a}_{j}^{\dagger},\nonumber \\
\sigma_{y}^{j} & = & \frac{1}{i}\left(\hat{a}_{j}^{\dagger}-\hat{a}_{j}\right),\nonumber \\
\sigma_{z}^{j} & = & \hat{a}_{j}-\hat{a}_{j}^{\dagger}.
\end{eqnarray}
One can verify that, for instance, $\vert0\rangle+\vert1\rangle$
is an eigenstate of $\sigma_{x}$: 
\begin{equation}
\langle\Phi\vert\sigma_{x}\left(\vert0\rangle+\vert1\rangle\right)=\langle\Phi\vert\left(\vert0\rangle+\vert1\rangle+\vert2\rangle\right)=\langle\Phi\vert\left(\vert0\rangle+\vert1\rangle\right)\,.
\end{equation}

Just as in the previous subsection, substituting $\hat{\rho}=\vert\Phi\rangle\langle\Phi\vert$
into~(\ref{eq:P-from-rho}) and performing the substitution~(\ref{eq:P-variable-change}),
we get the positive-P function:
\begin{eqnarray}
P & = & \frac{1}{2\pi^{2M}}e^{-\vert\vec{\nu}\vert^{2}}e^{-\vert\vec{\mu}\vert^{2}}\nonumber \\
 &  & \times\left(1+\prod_{j=1}^{M}\mu_{j}^{*}\mu_{j}+e^{-i\phi}\prod_{j=1}^{M}\mu_{j}^{*}+e^{i\phi}\prod_{j=1}^{M}\mu_{j}\right)\nonumber \\
 & = & \frac{1}{2\pi^{2M}}e^{-\vert\vec{\nu}\vert^{2}}e^{-\vert\vec{\mu}\vert^{2}}\left|\prod_{j=1}^{M}\mu_{j}+e^{-i\phi}\right|^{2}.
\end{eqnarray}

The target distribution can be sampled using von Neumann rejection
sampling:
\begin{equation}
P\le2G(\vec{\nu})P_{0}(\vec{\mu}),
\end{equation}
where the reference distributions are now:
\begin{equation}
G(\vec{\nu})=\frac{1}{\pi^{2M}}e^{-\vert\vec{\nu}\vert^{2}},
\end{equation}

\begin{equation}
P_{0}=\frac{1}{2\pi^{M}}e^{-\vert\vec{\mu}\vert^{2}}\left(\left|\prod_{j=1}^{M}\mu_{j}\right|^{2}+1\right).
\end{equation}

In this representation the expectation of the target operator is:
\begin{eqnarray}
\langle\hat{A}\rangle & = & \int d^{2M}\vec{\alpha}d^{2M}\vec{\beta}P\left(\vec{\mu}(\vec{\alpha},\vec{\beta}),\vec{\nu}(\vec{\alpha},\vec{\beta})\right)\nonumber \\
 &  & \times\prod_{j=1}^{M}\left(\left(\alpha_{j}+is_{j}\beta_{j}\right)e^{-is_{j}\theta_{j}}\right).
\end{eqnarray}

In Figure (\ref{fig:Spin-moments-P}), we show the distribution of
results with the positive-P representations, for a portion of the
Ardehali inequality for the case $M=2$, given by:
\begin{equation}
F_{XY}=-\langle\hat{\sigma}_{1}^{x}\hat{\sigma}_{2}^{x}\rangle+\langle\hat{\sigma}_{1}^{y}\hat{\sigma}_{2}^{y}\rangle.\label{eq:F-2-particles-1}
\end{equation}
In a LHV theory the values of $\mathrm{Re}\,\sigma_{x}^{1}$ and $\mathrm{Re}\,\sigma_{x}^{2}$
are limited to the range $[-1,1]$, but clearly our results are not
limited to that range. This essential feature means that Bell's theorem
does not limit our results, because the sampled values are not the
same as their physical eigenvalues~\cite{Reid1986-violations}. The
connection with weak values~\cite{Aharonov1988}\textcolor{red}{{}
}has been discussed in a previous paper~\cite{Rosales-Zarate2014}.
This demonstrates an essential feature of this phase-space representation:
it is analogous to a weak-value measurement, giving results outside
the normal range of the eigenvalues.

We note that this is still a positive phase-space representation,
valid in a more limited subspace than before, but certainly able to
represent the GHZ state. However, both approaches have the drawback
that they use methods designed to represent infinite dimensional Hilbert
spaces, which is not a good match to the GHZ state requirements.

\section{\label{sec:Sampling-GHZ-Qfunction}Sampling GHZ states with the Q-function}

The Hilbert space occupied by the GHZ state is a finite-dimensional
Hilbert space for which such an infinite-dimensional bosonic mapping
is not strictly necessary. Next we turn to methods that are more suited
to the task of representing finite dimensional states. Our interest
in doing this is to determine if this can improve the sampling properties.

The Q-function for bosons was first introduced by Husimi~\cite{Husimi1940}
as an expectation value of the density matrix in an over-complete
coherent-state basis. It gives a mapping of a general many-body density
matrix into a unique, positive distribution. This method has been
widely used as a method to probabilistically represent statistical
properties in quantum optics. It has had a diverse range of applications,
mostly in tomography. 

The same technique can be used to define a multipartite Q-function
based on SU(2) coherent states, as an alternative and more efficient
means of phase-space sampling for qubits.

\subsection{The SU(2) Q-function}

For purposes of calculations, we will consider as the basis set an
un-normalized version of the SU(2) coherent states~\cite{Arecchi1972,Radcliffe1971,Zhang1990}
defined as:
\begin{equation}
\left\Vert \vec{z}\,\right\rangle =\prod_{j=1}^{M}\left(\left|0\right\rangle _{j}+z_{j}\left|1\right\rangle _{j}\right).\label{eq:un-normalized_SU2CS-1}
\end{equation}

In terms of this un-normalized state the resolution of unity is given
by:
\begin{equation}
\int d^{2}\vec{z}\left(\prod_{j=1}^{M}{\cal N}\left(\left|z_{j}\right|^{2}\right)\right)\left\Vert \vec{z}\,\right\rangle \left\langle \vec{z}\,\right\Vert =\hat{1}.\label{eq:un-normResUnity_SU2CS-1}
\end{equation}
Here we have defined the normalization factor ${\cal N}\left(\left|z_{j}\right|^{2}\right)$
as:
\begin{equation}
{\cal N}\left(\left|z_{j}\right|^{2}\right)=\frac{2}{\pi}\frac{1}{\left(1+\left|z_{j}\right|^{2}\right)^{3}}.
\end{equation}
Using the resolution of unity for the un-normalized SU(2) coherent
states~(\ref{eq:un-normResUnity_SU2CS-1}), we can define a Q-function:
\begin{equation}
Q(\vec{z})=\left[\prod_{j=1}^{M}{\cal N}\left(\left|z_{j}\right|^{2}\right)\right]\left\langle \vec{z}\,\right\Vert \hat{\rho}\left\Vert \vec{z}\,\right\rangle ,\label{eq:Qf_un-norm_SU2CS-1}
\end{equation}
which has the property that:
\begin{equation}
\int d^{2}\vec{z}Q(\vec{z})=1.
\end{equation}
This Q-function is positive definite and is defined for any quantum
density matrix and is normalized to one. 

In our GHZ state of interest~(\ref{eq:ghz-full}) we denote $\left|\downarrow\right\rangle =\left|0\right\rangle $
and $\left|\uparrow\right\rangle =\left|1\right\rangle $. Hence for
the density matrix $\hat{\rho}=\vert\Phi\rangle\langle\Phi\vert$
we obtain:
\begin{eqnarray}
\left\langle \vec{z}\,\right\Vert \hat{\rho}\left\Vert \vec{z}\,\right\rangle  & = & \left\langle \vec{z}\,\right\Vert \vert\Phi\rangle\langle\Phi\vert\left\Vert \vec{z}\,\right\rangle =\left|\left\langle \vec{z}\,\right\Vert \vert\Phi\rangle\right|^{2}\nonumber \\
 & = & \frac{1}{2}\left|\prod_{j}\left(_{j}\left\langle 0\right|+z_{j}^{*}\,_{j}\left\langle 1\right|\right)\left(\vert1\ldots1\rangle+e^{i\phi}\vert0\ldots0\rangle\right)\right|^{2}\nonumber \\
 & = & \frac{1}{2}\left|\prod_{j}z_{j}+e^{-i\phi}\right|^{2}.
\end{eqnarray}
Therefore the Q-function for our states of interest is:
\begin{eqnarray}
Q(\vec{z}) & = & \frac{1}{2}\left(\frac{2}{\pi}\right)^{M}\prod_{j=1}^{M}\frac{1}{\left(1+\left|z_{j}\right|^{2}\right)^{3}}\left|\prod_{j}z_{j}+e^{-i\phi}\right|^{2}.
\end{eqnarray}

The expectation value of $\hat{A}$ can be expressed in terms of the
Q-function using~(\ref{eq:Q-spin-moments}), the details of the evaluations
are shown in the next section, and the fact that $\sigma_{x}^{j}=2\hat{S}_{x}^{j}=\hat{S}_{+}^{j}+\hat{S}_{-}^{j}$
and $\sigma_{y}^{j}=2\hat{S}_{y}^{j}=(\hat{S}_{+}^{j}-\hat{S}_{-}^{j})/i$,
hence:
\begin{equation}
\hat{A}=\prod_{j=1}^{M}\left(\left((1+s_{j})\hat{S}_{+}^{j}+(1-s_{j})\hat{S}_{-}^{j}\right)e^{-is_{j}\theta_{j}}\right).
\end{equation}
Therefore the expectation value of the target operator using the Q-function
is:
\begin{eqnarray}
\langle\hat{A}\rangle & = & \langle\Phi\vert\hat{A}\vert\Phi\rangle\nonumber \\
 & = & \int d\vec{z}Q(\vec{z})\frac{3^{M}}{\prod_{j}\left(1+\left|z_{j}\right|^{2}\right)}\\
 &  & \times\prod_{j=1}^{M}\left(\left((1+s_{j})z_{j}^{*}+(1-s_{j})z_{j}\right)e^{-is_{j}\theta_{j}}\right).\nonumber 
\end{eqnarray}

\subsection{Evaluation of moments\label{sec:Appendix_EvaluationMoments-1}}

In this section we show the evaluation of the moments of the form
$\left\langle \prod_{j}\hat{S}_{d_{j}}^{j}\right\rangle $ with directions
$d_{j}\in\left\{ -,+\right\} $, in terms of the SU(2) Q-function.
In order to evaluate the moments, we notice that we can express the
action of the raising spin operators on the un-normalized SU(2) coherent
state $\left\Vert \mathbf{z}\right\rangle $ as a derivative of the
SU(2) coherent state $\left\Vert \vec{z}\right\rangle $, so that:
\begin{eqnarray}
\hat{S}_{+}^{j}\left\Vert \vec{z}\right\rangle  & = & \hat{S}_{+}^{j}\left(\prod_{j}e^{\hat{S}_{+}^{j}z_{j}}\left|0\right\rangle _{j}\right)\nonumber \\
 & = & \frac{\partial}{\partial z_{j}}\left(\prod_{j}e^{\hat{S}_{+}^{j}z_{j}}\left|0\right\rangle _{j}\right)\nonumber \\
 & = & \frac{\partial}{\partial z_{j}}\left\Vert \vec{z}\right\rangle .
\end{eqnarray}

Similarly, there is a conjugate expression: 
\begin{eqnarray}
\left\langle \vec{z}\right\Vert \hat{S}_{-}^{j} & = & \left(\left\langle 0\right|e^{\hat{\bm{S}}_{-}\cdot\vec{z}}\right)\hat{S}_{-}^{j}\nonumber \\
 & = & \frac{\partial}{\partial z_{j}^{*}}\left(\prod_{j}\,_{j}\left\langle 0\right|e^{\hat{S}_{-}^{j}z_{j}}\right)\nonumber \\
 & = & \frac{\partial}{\partial z_{j}^{*}}\left\langle \vec{z}\right\Vert ,
\end{eqnarray}
while for the $z$-direction one obtains: 
\begin{eqnarray}
\hat{S}_{z}^{j}\left\Vert \mathbf{z}\right\rangle  & = & \hat{S}_{z}^{j}\left(\prod_{j}e^{\hat{S}_{+}^{j}z_{j}}\left|0\right\rangle _{j}\right)\nonumber \\
 & = & \prod_{j}\frac{1}{2}\left(-\left|0\right\rangle _{j}+z_{j}\left|1\right\rangle _{j}\right)\nonumber \\
 & = & \prod_{j}\frac{1}{2}\left(2z_{j}\frac{\partial}{\partial z_{j}}-1\right)\left\Vert \mathbf{z}\right\rangle .
\end{eqnarray}
Here we have used that:
\begin{eqnarray}
z_{j}\frac{\partial}{\partial z_{j}}\left(\left|0\right\rangle _{j}+z_{j}\left|1\right\rangle _{j}\right) & = & z_{j}\left|1\right\rangle _{j},
\end{eqnarray}
and hence the last identity above is obtained from:
\begin{eqnarray}
2z_{j}\frac{\partial}{\partial z_{j}}\left\Vert \mathbf{z}\right\rangle -\left\Vert \mathbf{z}\right\rangle  & = & z_{j}\left|1\right\rangle _{j}-\left|0\right\rangle _{j}.
\end{eqnarray}

Next, we evaluate the moments of the spin operators $\hat{S}_{+}^{j}$,
$\hat{S}_{-}^{j}$ and $\hat{S}_{z}^{j}$ using the resolution of
unity~(\ref{eq:un-normResUnity_SU2CS-1}) as well as the definition
of the Q-function~(\ref{eq:Qf_un-norm_SU2CS-1}) so that:
\begin{eqnarray}
\left\langle \prod_{j}\hat{S}_{d_{j}}^{j}\right\rangle  & = & {\rm Tr}\left[\hat{\rho}\prod_{j}\hat{S}_{d_{j}}^{j}\right]\\
 & = & \int d^{2}\vec{z}\prod_{j}{\cal N}\left(\left|z_{j}\right|^{2}\right)\nonumber \\
 &  & \times\left(\prod_{d_{j}=-}\frac{\partial}{\partial z_{j}^{*}}\right)\left\langle \vec{z}\right\Vert \hat{\rho}\left(\prod_{d_{j}=+}\frac{\partial}{\partial z_{j}}\right)\left\Vert \vec{z}\right\rangle \nonumber 
\end{eqnarray}
Integrating by parts for each $j$, providing that the boundary terms
vanish, we get:
\begin{eqnarray}
\left\langle \prod_{j}\hat{S}_{d_{j}}^{j}\right\rangle  & = & (-1)^{M}\int d^{2}\vec{z}\left\langle \vec{z}\right\Vert \hat{\rho}\left\Vert \vec{z}\right\rangle \label{eq:Q-spin-moments}\\
 &  & \times\prod_{j,d_{j}=-}\frac{\partial{\cal N}\left(\left|z_{j}\right|^{2}\right)}{\partial z_{j}^{*}}\prod_{j,d_{j}=+}\frac{\partial{\cal N}\left(\left|z_{j}\right|^{2}\right)}{\partial z_{j}}\nonumber \\
 & = & \int d^{2}\vec{z}Q(\vec{z})\frac{3^{M}}{\prod_{j}\left(1+\left|z_{j}\right|^{2}\right)}\prod_{j,d_{j}=-}z_{j}\prod_{j,d_{j}=+}z_{j}^{*}.\nonumber \\
 &  & \,\nonumber 
\end{eqnarray}
Here we have used that the derivative of ${\cal N}\left(\left|z_{j}\right|^{2}\right)$
is:
\begin{eqnarray}
\frac{1}{{\cal N}\left(\left|z_{j}\right|^{2}\right)}\frac{\partial}{\partial z_{j}}{\cal N}\left(\left|z_{j}\right|^{2}\right) & = & \frac{2}{\pi{\cal N}\left(\left|z_{j}\right|^{2}\right)}\frac{\partial}{\partial z_{j}}\frac{1}{\left(1+\left|z_{j}\right|^{2}\right)^{3}}\nonumber \\
 & = & \frac{-3z_{j}^{*}}{\left(1+\left|z_{j}\right|^{2}\right)}.\label{eq:derivative_Norm_nmode-1}\\
 &  & \,\nonumber 
\end{eqnarray}

Here, the results of Q-function sampling of the GHZ state are presented.
Firstly we show the results of the difference between the calculations
with the positive-P and SU(2)-Q representations, using a portion of
the Ardehali inequality for the case $M=2$, given as previously by:
\begin{equation}
F_{XY}=-\langle\hat{\sigma}_{1}^{x}\hat{\sigma}_{2}^{x}\rangle+\langle\hat{\sigma}_{1}^{y}\hat{\sigma}_{2}^{y}\rangle.\label{eq:F-2-particles}
\end{equation}

\begin{figure}
\includegraphics{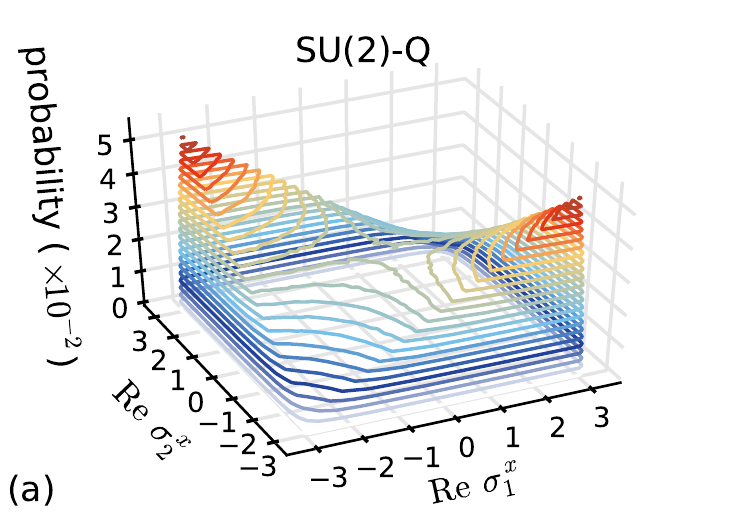}\\
\includegraphics{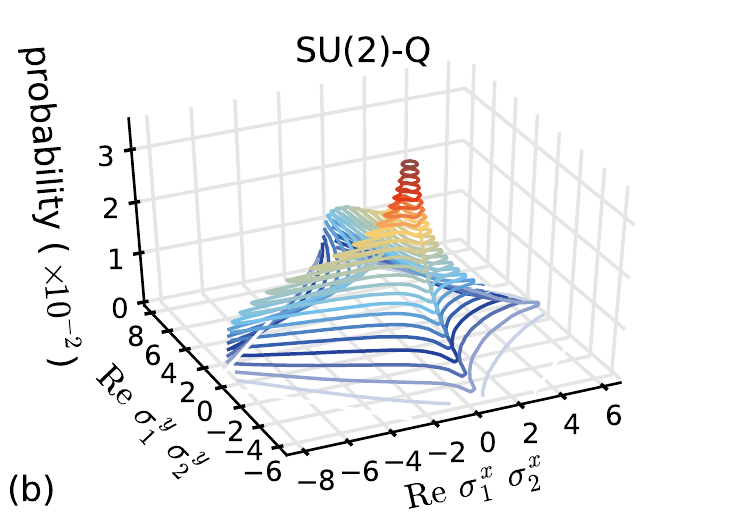}

\protect\caption{Correlations for the different parts of the quantity~(\ref{eq:F-2-particles})
in the SU(2)-Q representation, $2^{26}$ samples.\label{fig:Spin-moments-Q}}
\end{figure}

In Fig.~\ref{fig:Spin-moments-Q}(a) we show the correlation between
the real parts of $\hat{\sigma}_{i}^{x}$, $i=1,\:2$ of the quantity
$F_{XY}$, while the correlation between the two terms of~(\ref{eq:F-2-particles})
is plotted in Fig.~~\ref{fig:Spin-moments-Q}(b). Once again, this
method is analogous to a weak-value measurement, giving results outside
the normal range of the eigenvalues.

For the spin operator $\hat{S}_{z}^{j}$ we obtain:
\begin{eqnarray}
\left\langle \prod_{j}\hat{S}_{z}^{j}\right\rangle  & = & {\rm Tr}\left[\prod_{j}\hat{\rho}\hat{S}_{z}^{j}\right]\\
 & = & \prod_{j}\int d^{2}z_{j}{\cal N}\left(\left|z_{j}\right|^{2}\right)\left\langle \mathbf{z}\right\Vert \hat{\rho}\left(z_{j}\frac{\partial}{\partial z}-\frac{1}{2}\right)\left\Vert \mathbf{z}\right\rangle \,.\nonumber 
\end{eqnarray}
Next we use the following result, which also involves partial integration:
\begin{eqnarray}
 &  & \prod_{j}\int d^{2}z_{j}{\cal N}\left(\left|z_{j}\right|^{2}\right)\left\langle \mathbf{z}\right\Vert \hat{\rho}z_{j}\frac{\partial}{\partial z}\left\Vert \mathbf{z}\right\rangle \nonumber \\
 &  & =-\prod_{j}\int d^{2}z_{j}\left\langle \mathbf{z}\right\Vert \hat{\rho}\left\Vert \mathbf{z}\right\rangle \frac{\partial}{\partial z_{j}}\left(z_{j}{\cal N}\left(\left|z_{j}\right|^{2}\right)\right)\nonumber \\
 &  & =\prod_{j}\int d^{2}z_{j}Q\left(\mathbf{z}\right)\left(\frac{3\left|z_{j}\right|^{2}}{\left(1+\left|z_{j}\right|^{2}\right)}-1\right)\,,
\end{eqnarray}
and leads to our final spin operator identity, 
\begin{equation}
\left\langle \prod_{j}\hat{S}_{z}^{j}\right\rangle =\prod_{j}\frac{3}{2}\int d^{2}z_{j}Q\left(\mathbf{z}\right)\left(\frac{\left|z_{j}\right|^{2}-1}{\left|z_{j}\right|^{2}+1}\right)\,.
\end{equation}

\section{\label{sec:Results}Multipartite Bell violation results}

To simulate multipartite Bell violations, the GHZ state~(\ref{eq:ghz-full})
was sampled using probabilistic random number generators using both
Q-function and positive P-distribution methods. Of the two positive
P-distribution mappings, the Schwinger representation method is less
compact, and has a larger sampling error. For the results graphed
here, we therefore chose the number state positive-P distribution.
Although more efficient than the Schwinger representation, this still
has a large basis set that corresponds to an infinite dimensional
Hilbert space, with a much larger dimension than is needed for the
GHZ state. 

The lowest sampling errors were obtained with the $SU(2)$ Q-distribution
method, which uses a much more compact Hilbert space, having a dimension
equal to the physical qubit dimension.

\subsection{Multipartite sampling error properties}

We initially investigate the scaling properties of the sampling errors
as the number of qubits $M$ is varied. This also determines the time
taken for the simulation to reach a predetermined error, since one
can include more parallel samples to reduce the simulated errors to
any desired level.

\begin{figure}
\includegraphics{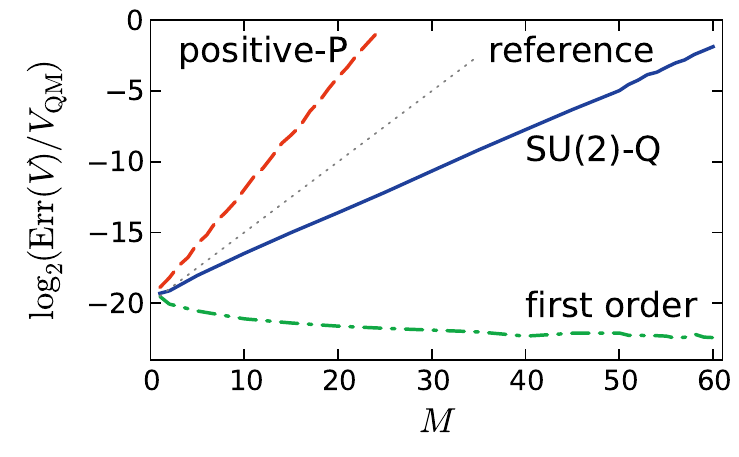}\protect\caption{Scaling properties for sampled correlations of multi-particle GHZ
states. Relative errors are plotted for high order ($V$) correlations,
(blue line) and first order correlations, or total number of ``spin-ups''
(green dashed line) using the SU(2)-Q representation with $2^{40}$
samples. The dotted reference line shows the point at which the sampling
errors would give scaling properties of an experimental measurement.
The red dotted line shows the scaling of the $V$ correlations using
the less efficient positive-P representation.\label{fig:GHZ-violations-1}}
\end{figure}

First we consider the scaling with system-size of the sampling errors
for\emph{ single} measurements of a low-order spin correlation (Fig.~\ref{fig:GHZ-violations-1}).
For low-order correlation we have chosen the total number of ``spin-ups''
$N=\langle\sum_{j=1}^{M}\left(\hat{\sigma}_{z}^{j}+1\right)/2\rangle$.
In this case we noticed that the sampling errors \emph{decreases}
as $M$ increases. 

In contrast to this, high-order correlations showed exponentially
increasing sampling error. The relative error in $V$ scales as $2^{M/3}$,
meaning that the time taken at constant error scales as $2^{2M/3}$.
This means that probabilistic sampling scales more favorably than
experiment, which would take time in proportion to $2^{M}$. Therefore,
the sampling takes place in times that scale $2^{M/3}$ times faster
than any possible experiment.

In practical terms, such laboratory measurements would be highly nontrivial,
due to the need to eliminate background noise for high-order correlations.
No correlation measurements of this size have been reported to date.
Experimentally, it is possible that such high-order correlations will
be reported in future. 

Even then, it is likely that one may only be able to measure a subset
of all the high-order correlations possible for large $M$ values.
This is because of the enormous time required to make all possible
correlation measurements for these inequalities, which is exponentially
slower than the phase-space simulation.

\subsection{Simulations of multipartite genuine Bell violations}

\begin{figure}
\includegraphics{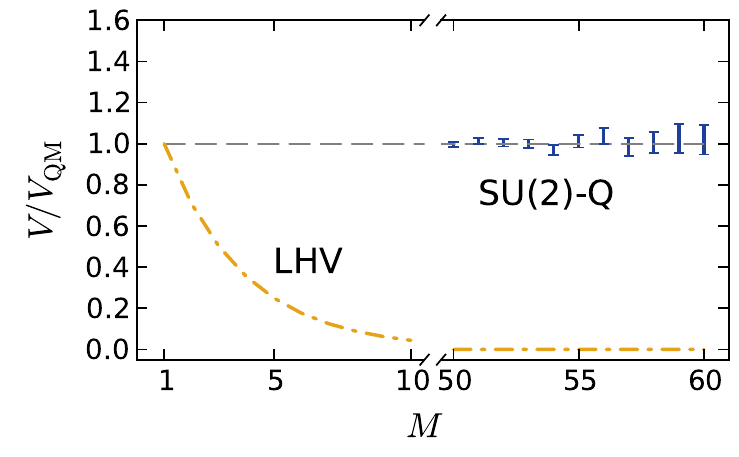}\\
\protect\caption{Violations for multi-particle GHZ states. Simulated Mermin violation
using SU(2)-Q representation with $2^{43}$ samples. The values of
expectations and errors are normalized by the quantum mechanical prediction
for the corresponding $M$. The horizontal grey dashed line gives
the quantum prediction. The error bars show the sampled result and
estimated sampling errors at each value of $M$. The dash-dotted line
is the LHV prediction, which gives a Bell violation when above this
line. Genuine multipartite Bell violations occur for even $M$ when
$V/V_{\mathrm{QM}}>1/\sqrt{2}$.\label{fig:GHZ-violations}}
\end{figure}

In Fig.~\ref{fig:GHZ-violations} we show the expectation value of
the multipartite, multi-measurement quantity $V$ compared with the
quantum mechanical prediction $\left\langle V\right\rangle _{QM}$
from sampling the $SU(2)$ Q-distribution. The dashed line is the
minimum correlation required to demonstrate a Bell violation, with
a number of qubits ranging from $M=2$ to $M=60$. For all cases we
verified clear Bell violations to at least $12$ standard deviations
from the classical limit. 

Genuine multipartite violations of LHV, requiring all $M$ observers
to participate, were verified for even $M$ to at least $4$ standard
deviations. These cases all satisfied the more stringent requirement
that: 
\begin{equation}
V/V_{\mathrm{QM}}>1/\sqrt{2}\,.
\end{equation}

The simulations were carried out using graphical processor unit (GPU)
technology at a clock speed of $1.2$ GHz, which allowed calculations
with $50$ GPUs on $22,000$ parallel computational cores. The plotted
results correspond in the $60$ qubit case to simulating the results
of a quintillion ($10^{18}$) distinct sixtieth order correlation
functions. This took less than $48$ hours. A reasonable estimate
of the laboratory time-scale for carrying out all possible correlation
measurements, at $10^{-3}s$ per measurement setting, is $3\times10^{7}$
years. This is more than $10^{9}$ times slower than the simulations.

\section{\label{sec:Decoherence-model}Decoherence simulations }

We have shown that it is possible to simulate genuine Bell violations,
as well as obtaining scaling laws for GHZ states using phase space
methods. But we can also ask whether it is also possible to use the
positive phase-space methods to simulate decoherence processes? In
order to answer this question, here we will focus on the question
of the study of dynamical noise and decoherence in ion traps, which
is an important issue in the observation of mesoscopic quantum effects~\cite{Brune1996}.
Ion traps have been widely used in order to create entangled states
and also to investigate the decay rate of GHZ states~\cite{Monz2011}. 

\begin{figure}
\includegraphics{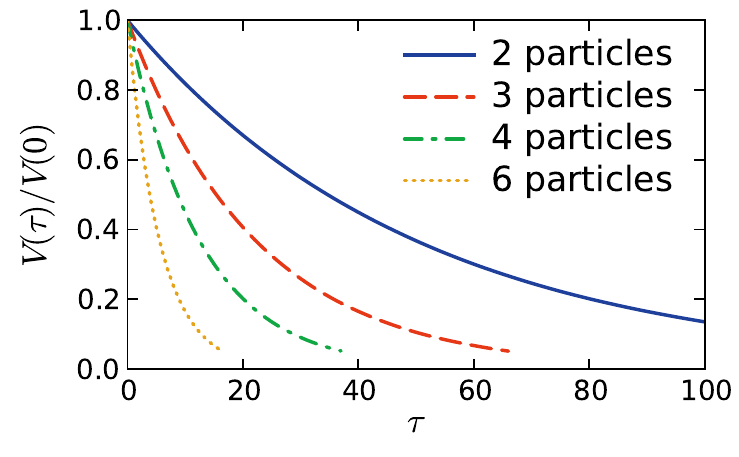}\protect\caption{Decay of the sampled quantity $V$ using the model of super-decoherence
of Eq~(\ref{eq:H-decoherence}), for 2 (solid blue line), 3 (red
dashed line), 4 (green dash-dotted line) and 6 (yellow dotted line)
particles, with decoherence rate $\epsilon=0.1$. The horizontal axis
is the dimensionless time, $\tau=t/\Delta t$.\label{fig:Decoherence}}
\end{figure}

Here we will follow the noise model of Monz et al.~\cite{Monz2011},
which was used to explain the observed super-decoherence found in
ion-trap experiments. This is physically due to the fact that the
magnetic field noise reservoir is correlated over all the qubits.
As a result, they do not decohere with independent noise or error
sources, as is often assumed theoretically. To model this, we assume
a delta-correlated magnetic field noise which is shared by all the
ions, such that 
\begin{equation}
\left\langle \Delta B(t)\Delta B(t')\right\rangle =\Delta B_{0}^{2}\delta(t-t')\,.
\end{equation}
In this case we assume that the interaction or noise Hamiltonian is:
\begin{equation}
\hat{H}=\frac{\mu\Delta B(t)}{2}\sum_{j=1}^{M}\hat{\sigma}_{z}^{j}.\label{eq:H-decoherence}
\end{equation}
This model can be simulated dynamically multiplying, in each of the
samples after every time step $\Delta t$, an independent noise term
$\exp\left(i\epsilon N\zeta_{j}\right)$ by the value corresponding
to the operator $\prod_{j}^{M}\left(\hat{\sigma}_{x}^{j}+\hat{\sigma}_{y}^{j}\right)$.
We use the respectively measurement choice $V$ of~(\ref{eq:V_Op})
for odd $M$ and even $M$. Here
\begin{equation}
\epsilon=\mu\Delta B_{0}\sqrt{\Delta t}/\hbar
\end{equation}
 defines the speed of the decoherence, and $\zeta_{j}$ is a Gaussian
random number such that
\begin{equation}
\left\langle \zeta_{j}\zeta_{j'}\right\rangle =\delta_{jj'}\,.
\end{equation}
 The results of the simulation are shown in Fig.~\ref{fig:Decoherence}.
This demonstrates the experimentally observed quadratic decoherence,
with decay times scaling with $1/M^{2}$ as $M$ increases over a
range comparable to current experiments, therefore showing the effect
of super-decoherence.

\section{\label{sec:Conclusions}Conclusions}

Our main result is that it is possible to sample quantum events probabilistically,
even when they display macroscopic quantum paradoxes and Schr\"odinger
cat behavior. This is not prohibited by the Bell inequality, although
we use standard digital computers. Our calculations generate a distribution
equivalent to the observables predicted by quantum mechanics. These
results demonstrate the potential for phase-space methods to simulate
macroscopic quantum superpositions.

We have demonstrated genuine multipartite Bell inequalities with up
to $60$ qubits. In all cases we have shown violations of these inequalities
using positive phase space distributions. We interested in the question
of whether or not probabilistic sampling can be carried out for GHZ
states. This is not obvious a-priori, since one might expect highly
nonclassical states to be hard to sample probabilistically. We also
have performed dynamical simulations of super-decoherence. 

These results demonstrate that the simulation of both low and high
order correlations is feasible, despite Bell violations. Some reasonable
conclusions about the advantages and limitations of these methods
are as follows. Probabilistic phase-space algorithms appear well suited
to low order correlations, including fundamentally nonclassical low
order Bell inequality violations. Higher order correlations generate
larger sampling errors with a probabilistic approach. 

We also find a classical parallelism which gives an an unexpected
exponential speed-up for qubit sampling, when calculating all the
high order correlations required for multipartite Bell violations.
Here the speed-up is relative to the corresponding experimental times,
and is exponential in the qubit number. This uniquely useful feature
of probabilistic phase-space methods is due to their ability to simultaneously
calculate many non-commuting observables in parallel. 

Such classical \emph{measurement} parallelism is complementary to
the \emph{state} parallelism of quantum mechanics. We have utilized
this in the calculation of the MABK function, to indicate genuine
violation of multipartite Bell inequalities. 
\begin{acknowledgments}
We wish to acknowledge research funding from the Australian Research
Council, as well as useful discussions with P. Deuar, R. Blatt and
B. Lanyon.
\end{acknowledgments}
\bibliography{MultiQsims}

\begin{thebibliography}{74}%
\makeatletter
\providecommand \@ifxundefined [1]{%
 \@ifx{#1\undefined}
}%
\providecommand \@ifnum [1]{%
 \ifnum #1\expandafter \@firstoftwo
 \else \expandafter \@secondoftwo
 \fi
}%
\providecommand \@ifx [1]{%
 \ifx #1\expandafter \@firstoftwo
 \else \expandafter \@secondoftwo
 \fi
}%
\providecommand \natexlab [1]{#1}%
\providecommand \enquote  [1]{``#1''}%
\providecommand \bibnamefont  [1]{#1}%
\providecommand \bibfnamefont [1]{#1}%
\providecommand \citenamefont [1]{#1}%
\providecommand \href@noop [0]{\@secondoftwo}%
\providecommand \href [0]{\begingroup \@sanitize@url \@href}%
\providecommand \@href[1]{\@@startlink{#1}\@@href}%
\providecommand \@@href[1]{\endgroup#1\@@endlink}%
\providecommand \@sanitize@url [0]{\catcode `\\12\catcode `\$12\catcode
  `\&12\catcode `\#12\catcode `\^12\catcode `\_12\catcode `\%12\relax}%
\providecommand \@@startlink[1]{}%
\providecommand \@@endlink[0]{}%
\providecommand \url  [0]{\begingroup\@sanitize@url \@url }%
\providecommand \@url [1]{\endgroup\@href {#1}{\urlprefix }}%
\providecommand \urlprefix  [0]{URL }%
\providecommand \Eprint [0]{\href }%
\providecommand \doibase [0]{http://dx.doi.org/}%
\providecommand \selectlanguage [0]{\@gobble}%
\providecommand \bibinfo  [0]{\@secondoftwo}%
\providecommand \bibfield  [0]{\@secondoftwo}%
\providecommand \translation [1]{[#1]}%
\providecommand \BibitemOpen [0]{}%
\providecommand \bibitemStop [0]{}%
\providecommand \bibitemNoStop [0]{.\EOS\space}%
\providecommand \EOS [0]{\spacefactor3000\relax}%
\providecommand \BibitemShut  [1]{\csname bibitem#1\endcsname}%
\let\auto@bib@innerbib\@empty
\bibitem [{\citenamefont {Dirac}(1929)}]{Dirac1929}%
  \BibitemOpen
  \bibfield  {author} {\bibinfo {author} {\bibfnamefont {P.~A.~M.}\
  \bibnamefont {Dirac}},\ }\href {\doibase 10.1098/rspa.1929.0094} {\bibfield
  {journal} {\bibinfo  {journal} {P. Roy. Soc. A.}\ }\textbf {\bibinfo {volume}
  {123}},\ \bibinfo {pages} {714} (\bibinfo {year} {1929})}\BibitemShut
  {NoStop}%
\bibitem [{\citenamefont {Feynman}(1982)}]{Feynman1982}%
  \BibitemOpen
  \bibfield  {author} {\bibinfo {author} {\bibfnamefont {R.~P.}\ \bibnamefont
  {Feynman}},\ }\href {\doibase 10.1007/BF02650179} {\bibfield  {journal}
  {\bibinfo  {journal} {Int. J. Theor. Phys.}\ }\textbf {\bibinfo {volume}
  {21}},\ \bibinfo {pages} {467} (\bibinfo {year} {1982})}\BibitemShut
  {NoStop}%
\bibitem [{\citenamefont {Haake}\ \emph {et~al.}(1979)\citenamefont {Haake},
  \citenamefont {King}, \citenamefont {Schr\"{o}der}, \citenamefont {Haus},\
  and\ \citenamefont {Glauber}}]{Haake1979}%
  \BibitemOpen
  \bibfield  {author} {\bibinfo {author} {\bibfnamefont {F.}~\bibnamefont
  {Haake}}, \bibinfo {author} {\bibfnamefont {H.}~\bibnamefont {King}},
  \bibinfo {author} {\bibfnamefont {G.}~\bibnamefont {Schr\"{o}der}}, \bibinfo
  {author} {\bibfnamefont {J.}~\bibnamefont {Haus}}, \ and\ \bibinfo {author}
  {\bibfnamefont {R.~J.}\ \bibnamefont {Glauber}},\ }\href {\doibase
  10.1103/PhysRevA.20.2047} {\bibfield  {journal} {\bibinfo  {journal} {Phys.
  Rev. A}\ }\textbf {\bibinfo {volume} {20}},\ \bibinfo {pages} {2047}
  (\bibinfo {year} {1979})}\BibitemShut {NoStop}%
\bibitem [{\citenamefont {Corney}\ \emph {et~al.}(2006)\citenamefont {Corney}
  \emph {et~al.}}]{Corney2006}%
  \BibitemOpen
  \bibfield  {author} {\bibinfo {author} {\bibfnamefont {J.~F.}\ \bibnamefont
  {Corney}} \emph {et~al.},\ }\href {\doibase 10.1103/PhysRevLett.97.023606}
  {\bibfield  {journal} {\bibinfo  {journal} {Phys. Rev. Lett.}\ }\textbf
  {\bibinfo {volume} {97}},\ \bibinfo {pages} {023606} (\bibinfo {year}
  {2006})}\BibitemShut {NoStop}%
\bibitem [{\citenamefont {Deuar}\ and\ \citenamefont
  {Drummond}(2007)}]{Deuar2007}%
  \BibitemOpen
  \bibfield  {author} {\bibinfo {author} {\bibfnamefont {P.~P.}\ \bibnamefont
  {Deuar}}\ and\ \bibinfo {author} {\bibfnamefont {P.~D.}\ \bibnamefont
  {Drummond}},\ }\href {\doibase 10.1103/PhysRevLett.98.120402} {\bibfield
  {journal} {\bibinfo  {journal} {Phys. Rev. Lett.}\ }\textbf {\bibinfo
  {volume} {98}},\ \bibinfo {pages} {120402} (\bibinfo {year}
  {2007})}\BibitemShut {NoStop}%
\bibitem [{\citenamefont {Alon}\ \emph {et~al.}(2008)\citenamefont {Alon},
  \citenamefont {Streltsov},\ and\ \citenamefont {Cederbaum}}]{Alon2008}%
  \BibitemOpen
  \bibfield  {author} {\bibinfo {author} {\bibfnamefont {O.~E.}\ \bibnamefont
  {Alon}}, \bibinfo {author} {\bibfnamefont {A.~I.}\ \bibnamefont {Streltsov}},
  \ and\ \bibinfo {author} {\bibfnamefont {L.~S.}\ \bibnamefont {Cederbaum}},\
  }\href {\doibase 10.1103/PhysRevA.77.033613} {\bibfield  {journal} {\bibinfo
  {journal} {Phys. Rev. A}\ }\textbf {\bibinfo {volume} {77}},\ \bibinfo
  {pages} {033613} (\bibinfo {year} {2008})}\BibitemShut {NoStop}%
\bibitem [{\citenamefont {Gambetta}\ \emph {et~al.}(2008)\citenamefont
  {Gambetta} \emph {et~al.}}]{Gambetta2008}%
  \BibitemOpen
  \bibfield  {author} {\bibinfo {author} {\bibfnamefont {J.}~\bibnamefont
  {Gambetta}} \emph {et~al.},\ }\href {\doibase 10.1103/PhysRevA.77.012112}
  {\bibfield  {journal} {\bibinfo  {journal} {Phys. Rev. A}\ }\textbf {\bibinfo
  {volume} {77}},\ \bibinfo {pages} {012112} (\bibinfo {year}
  {2008})}\BibitemShut {NoStop}%
\bibitem [{\citenamefont {Trotzky}\ \emph {et~al.}(2012)\citenamefont {Trotzky}
  \emph {et~al.}}]{Trotzky2012}%
  \BibitemOpen
  \bibfield  {author} {\bibinfo {author} {\bibfnamefont {S.}~\bibnamefont
  {Trotzky}} \emph {et~al.},\ }\href {\doibase 10.1038/nphys2232} {\bibfield
  {journal} {\bibinfo  {journal} {Nat. Phys.}\ }\textbf {\bibinfo {volume}
  {8}},\ \bibinfo {pages} {325} (\bibinfo {year} {2012})}\BibitemShut {NoStop}%
\bibitem [{\citenamefont {Cirac}\ and\ \citenamefont
  {Zoller}(2003)}]{Cirac2003}%
  \BibitemOpen
  \bibfield  {author} {\bibinfo {author} {\bibfnamefont {J.~I.}\ \bibnamefont
  {Cirac}}\ and\ \bibinfo {author} {\bibfnamefont {P.}~\bibnamefont {Zoller}},\
  }\href {\doibase 10.1126/science.1085130} {\bibfield  {journal} {\bibinfo
  {journal} {Science}\ }\textbf {\bibinfo {volume} {301}},\ \bibinfo {pages}
  {176} (\bibinfo {year} {2003})}\BibitemShut {NoStop}%
\bibitem [{\citenamefont {Jaksch}\ and\ \citenamefont
  {Zoller}(2005)}]{Jaksch2005}%
  \BibitemOpen
  \bibfield  {author} {\bibinfo {author} {\bibfnamefont {D.}~\bibnamefont
  {Jaksch}}\ and\ \bibinfo {author} {\bibfnamefont {P.}~\bibnamefont
  {Zoller}},\ }\href {\doibase 10.1016/j.aop.2004.09.010} {\bibfield  {journal}
  {\bibinfo  {journal} {Ann. Phys.}\ }\textbf {\bibinfo {volume} {315}},\
  \bibinfo {pages} {52} (\bibinfo {year} {2005})}\BibitemShut {NoStop}%
\bibitem [{\citenamefont {Buluta}\ and\ \citenamefont
  {Nori}(2009)}]{Buluta2009}%
  \BibitemOpen
  \bibfield  {author} {\bibinfo {author} {\bibfnamefont {I.}~\bibnamefont
  {Buluta}}\ and\ \bibinfo {author} {\bibfnamefont {F.}~\bibnamefont {Nori}},\
  }\href {\doibase 10.1126/science.1177838} {\bibfield  {journal} {\bibinfo
  {journal} {Science}\ }\textbf {\bibinfo {volume} {326}},\ \bibinfo {pages}
  {108} (\bibinfo {year} {2009})}\BibitemShut {NoStop}%
\bibitem [{\citenamefont {Islam}\ \emph {et~al.}(2011)\citenamefont {Islam}
  \emph {et~al.}}]{Islam2011}%
  \BibitemOpen
  \bibfield  {author} {\bibinfo {author} {\bibfnamefont {R.}~\bibnamefont
  {Islam}} \emph {et~al.},\ }\href {\doibase 10.1038/ncomms1374} {\bibfield
  {journal} {\bibinfo  {journal} {Nat. Commun.}\ }\textbf {\bibinfo {volume}
  {2}},\ \bibinfo {pages} {377} (\bibinfo {year} {2011})}\BibitemShut {NoStop}%
\bibitem [{\citenamefont {Georgescu}\ \emph {et~al.}(2014)\citenamefont
  {Georgescu}, \citenamefont {Ashhab},\ and\ \citenamefont
  {Nori}}]{Georgescu2014}%
  \BibitemOpen
  \bibfield  {author} {\bibinfo {author} {\bibfnamefont {I.~M.}\ \bibnamefont
  {Georgescu}}, \bibinfo {author} {\bibfnamefont {S.}~\bibnamefont {Ashhab}}, \
  and\ \bibinfo {author} {\bibfnamefont {F.}~\bibnamefont {Nori}},\ }\href
  {\doibase 10.1103/RevModPhys.86.153} {\bibfield  {journal} {\bibinfo
  {journal} {Rev. Mod. Phys.}\ }\textbf {\bibinfo {volume} {86}},\ \bibinfo
  {pages} {153} (\bibinfo {year} {2014})}\BibitemShut {NoStop}%
\bibitem [{\citenamefont {Lanyon}\ \emph {et~al.}(2011)\citenamefont {Lanyon}
  \emph {et~al.}}]{Lanyon2011}%
  \BibitemOpen
  \bibfield  {author} {\bibinfo {author} {\bibfnamefont {B.~P.}\ \bibnamefont
  {Lanyon}} \emph {et~al.},\ }\href {\doibase 10.1126/science.1208001}
  {\bibfield  {journal} {\bibinfo  {journal} {Science}\ }\textbf {\bibinfo
  {volume} {334}},\ \bibinfo {pages} {57} (\bibinfo {year} {2011})}\BibitemShut
  {NoStop}%
\bibitem [{\citenamefont {Glauber}\ and\ \citenamefont
  {Haake}(1978)}]{Glauber1978}%
  \BibitemOpen
  \bibfield  {author} {\bibinfo {author} {\bibfnamefont {R.~J.}\ \bibnamefont
  {Glauber}}\ and\ \bibinfo {author} {\bibfnamefont {F.}~\bibnamefont
  {Haake}},\ }\href {\doibase 10.1016/0375-9601(78)90747-8} {\bibfield
  {journal} {\bibinfo  {journal} {Phys. Lett. A}\ }\textbf {\bibinfo {volume}
  {68}},\ \bibinfo {pages} {29} (\bibinfo {year} {1978})}\BibitemShut {NoStop}%
\bibitem [{\citenamefont {Drummond}\ and\ \citenamefont
  {Gardiner}(1980)}]{Drummond1980}%
  \BibitemOpen
  \bibfield  {author} {\bibinfo {author} {\bibfnamefont {P.~D.}\ \bibnamefont
  {Drummond}}\ and\ \bibinfo {author} {\bibfnamefont {C.~W.}\ \bibnamefont
  {Gardiner}},\ }\href {\doibase 10.1088/0305-4470/13/7/018} {\bibfield
  {journal} {\bibinfo  {journal} {J. Phys. A: Math. Gen.}\ }\textbf {\bibinfo
  {volume} {13}},\ \bibinfo {pages} {2353} (\bibinfo {year}
  {1980})}\BibitemShut {NoStop}%
\bibitem [{\citenamefont {Carter}\ \emph {et~al.}(1987)\citenamefont {Carter},
  \citenamefont {Drummond}, \citenamefont {Reid},\ and\ \citenamefont
  {Shelby}}]{Carter1987}%
  \BibitemOpen
  \bibfield  {author} {\bibinfo {author} {\bibfnamefont {S.~J.}\ \bibnamefont
  {Carter}}, \bibinfo {author} {\bibfnamefont {P.~D.}\ \bibnamefont
  {Drummond}}, \bibinfo {author} {\bibfnamefont {M.~D.}\ \bibnamefont {Reid}},
  \ and\ \bibinfo {author} {\bibfnamefont {R.~M.}\ \bibnamefont {Shelby}},\
  }\href {\doibase 10.1103/PhysRevLett.58.1841} {\bibfield  {journal} {\bibinfo
   {journal} {Phys. Rev. Lett.}\ }\textbf {\bibinfo {volume} {58}},\ \bibinfo
  {pages} {1841} (\bibinfo {year} {1987})}\BibitemShut {NoStop}%
\bibitem [{\citenamefont {Rosenbluh}\ and\ \citenamefont
  {Shelby}(1991)}]{Rosenbluh1991}%
  \BibitemOpen
  \bibfield  {author} {\bibinfo {author} {\bibfnamefont {M.}~\bibnamefont
  {Rosenbluh}}\ and\ \bibinfo {author} {\bibfnamefont {R.~M.}\ \bibnamefont
  {Shelby}},\ }\href {\doibase 10.1103/PhysRevLett.66.153} {\bibfield
  {journal} {\bibinfo  {journal} {Phys. Rev. Lett.}\ }\textbf {\bibinfo
  {volume} {66}},\ \bibinfo {pages} {153} (\bibinfo {year} {1991})}\BibitemShut
  {NoStop}%
\bibitem [{\citenamefont {Drummond}\ \emph {et~al.}(1993)\citenamefont
  {Drummond}, \citenamefont {Shelby}, \citenamefont {Friberg},\ and\
  \citenamefont {Yamamoto}}]{Drummond1993-solitons}%
  \BibitemOpen
  \bibfield  {author} {\bibinfo {author} {\bibfnamefont {P.~D.}\ \bibnamefont
  {Drummond}}, \bibinfo {author} {\bibfnamefont {R.~M.}\ \bibnamefont
  {Shelby}}, \bibinfo {author} {\bibfnamefont {S.~R.}\ \bibnamefont {Friberg}},
  \ and\ \bibinfo {author} {\bibfnamefont {Y.}~\bibnamefont {Yamamoto}},\
  }\href {\doibase 10.1038/365307a0} {\bibfield  {journal} {\bibinfo  {journal}
  {Nature}\ }\textbf {\bibinfo {volume} {365}},\ \bibinfo {pages} {307}
  (\bibinfo {year} {1993})}\BibitemShut {NoStop}%
\bibitem [{\citenamefont {Heersink}\ \emph {et~al.}(2005)\citenamefont
  {Heersink}, \citenamefont {Josse}, \citenamefont {Leuchs},\ and\
  \citenamefont {Andersen}}]{Heersink2005}%
  \BibitemOpen
  \bibfield  {author} {\bibinfo {author} {\bibfnamefont {J.}~\bibnamefont
  {Heersink}}, \bibinfo {author} {\bibfnamefont {V.}~\bibnamefont {Josse}},
  \bibinfo {author} {\bibfnamefont {G.}~\bibnamefont {Leuchs}}, \ and\ \bibinfo
  {author} {\bibfnamefont {U.~L.}\ \bibnamefont {Andersen}},\ }\href {\doibase
  10.1364/OL.30.001192} {\bibfield  {journal} {\bibinfo  {journal} {Opt.
  Lett.}\ }\textbf {\bibinfo {volume} {30}},\ \bibinfo {pages} {1192} (\bibinfo
  {year} {2005})}\BibitemShut {NoStop}%
\bibitem [{\citenamefont {Lewis-Swan}\ and\ \citenamefont
  {Kheruntsyan}(2014)}]{Lewis-Swan2014}%
  \BibitemOpen
  \bibfield  {author} {\bibinfo {author} {\bibfnamefont {R.~J.}\ \bibnamefont
  {Lewis-Swan}}\ and\ \bibinfo {author} {\bibfnamefont {K.~V.}\ \bibnamefont
  {Kheruntsyan}},\ }\href {\doibase 10.1038/ncomms4752} {\bibfield  {journal}
  {\bibinfo  {journal} {Nat. Commun.}\ }\textbf {\bibinfo {volume} {5}},\
  \bibinfo {pages} {3752} (\bibinfo {year} {2014})}\BibitemShut {NoStop}%
\bibitem [{\citenamefont {Rosales-Z\'{a}rate}\ \emph
  {et~al.}(2014)\citenamefont {Rosales-Z\'{a}rate}, \citenamefont {Opanchuk},
  \citenamefont {Drummond},\ and\ \citenamefont {Reid}}]{Rosales-Zarate2014}%
  \BibitemOpen
  \bibfield  {author} {\bibinfo {author} {\bibfnamefont {L.}~\bibnamefont
  {Rosales-Z\'{a}rate}}, \bibinfo {author} {\bibfnamefont {B.}~\bibnamefont
  {Opanchuk}}, \bibinfo {author} {\bibfnamefont {P.~D.}\ \bibnamefont
  {Drummond}}, \ and\ \bibinfo {author} {\bibfnamefont {M.~D.}\ \bibnamefont
  {Reid}},\ }\href {http://arxiv.org/abs/1405.1168} {\  (\bibinfo {year}
  {2014})},\ \Eprint {http://arxiv.org/abs/1405.1168} {arXiv:1405.1168}
  \BibitemShut {NoStop}%
\bibitem [{\citenamefont {Greenberger}\ \emph {et~al.}(1989)\citenamefont
  {Greenberger}, \citenamefont {Horne},\ and\ \citenamefont
  {Zeilinger}}]{Greenberger1989}%
  \BibitemOpen
  \bibfield  {author} {\bibinfo {author} {\bibfnamefont {D.~M.}\ \bibnamefont
  {Greenberger}}, \bibinfo {author} {\bibfnamefont {M.~A.}\ \bibnamefont
  {Horne}}, \ and\ \bibinfo {author} {\bibfnamefont {A.}~\bibnamefont
  {Zeilinger}},\ }\href@noop {} {\emph {\bibinfo {title} {{Bell's Theorem,
  Quantum Theory and Conceptions of the Universe}}}},\ edited by\ \bibinfo
  {editor} {\bibfnamefont {M.}~\bibnamefont {Kafatos}}\ (\bibinfo  {publisher}
  {Springer},\ \bibinfo {year} {1989})\ p.\ \bibinfo {pages} {348}\BibitemShut
  {NoStop}%
\bibitem [{\citenamefont {Lu}\ \emph {et~al.}(2007)\citenamefont {Lu} \emph
  {et~al.}}]{Lu2007}%
  \BibitemOpen
  \bibfield  {author} {\bibinfo {author} {\bibfnamefont {C.-Y.}\ \bibnamefont
  {Lu}} \emph {et~al.},\ }\href {\doibase 10.1038/nphys507} {\bibfield
  {journal} {\bibinfo  {journal} {Nat. Phys.}\ }\textbf {\bibinfo {volume}
  {3}},\ \bibinfo {pages} {91} (\bibinfo {year} {2007})}\BibitemShut {NoStop}%
\bibitem [{\citenamefont {Leibfried}\ \emph {et~al.}(2005)\citenamefont
  {Leibfried} \emph {et~al.}}]{Leibfried2005-creation}%
  \BibitemOpen
  \bibfield  {author} {\bibinfo {author} {\bibfnamefont {D.}~\bibnamefont
  {Leibfried}} \emph {et~al.},\ }\href {\doibase 10.1038/nature04251}
  {\bibfield  {journal} {\bibinfo  {journal} {Nature}\ }\textbf {\bibinfo
  {volume} {438}},\ \bibinfo {pages} {639} (\bibinfo {year}
  {2005})}\BibitemShut {NoStop}%
\bibitem [{\citenamefont {Lanyon}\ \emph {et~al.}(2014)\citenamefont {Lanyon}
  \emph {et~al.}}]{Lanyon2014}%
  \BibitemOpen
  \bibfield  {author} {\bibinfo {author} {\bibfnamefont {B.~P.}\ \bibnamefont
  {Lanyon}} \emph {et~al.},\ }\href {\doibase 10.1103/PhysRevLett.112.100403}
  {\bibfield  {journal} {\bibinfo  {journal} {Phys. Rev. Lett.}\ }\textbf
  {\bibinfo {volume} {112}},\ \bibinfo {pages} {100403} (\bibinfo {year}
  {2014})}\BibitemShut {NoStop}%
\bibitem [{\citenamefont
  {Mermin}(1990{\natexlab{a}})}]{Mermin1990-entanglement}%
  \BibitemOpen
  \bibfield  {author} {\bibinfo {author} {\bibfnamefont {N.~D.}\ \bibnamefont
  {Mermin}},\ }\href {\doibase 10.1103/PhysRevLett.65.1838} {\bibfield
  {journal} {\bibinfo  {journal} {Phys. Rev. Lett.}\ }\textbf {\bibinfo
  {volume} {65}},\ \bibinfo {pages} {1838} (\bibinfo {year}
  {1990}{\natexlab{a}})}\BibitemShut {NoStop}%
\bibitem [{\citenamefont {Aharonov}\ \emph {et~al.}(1988)\citenamefont
  {Aharonov}, \citenamefont {Albert},\ and\ \citenamefont
  {Vaidman}}]{Aharonov1988}%
  \BibitemOpen
  \bibfield  {author} {\bibinfo {author} {\bibfnamefont {Y.}~\bibnamefont
  {Aharonov}}, \bibinfo {author} {\bibfnamefont {D.~Z.}\ \bibnamefont
  {Albert}}, \ and\ \bibinfo {author} {\bibfnamefont {L.}~\bibnamefont
  {Vaidman}},\ }\href {\doibase 10.1103/PhysRevLett.60.1351} {\bibfield
  {journal} {\bibinfo  {journal} {Phys. Rev. Lett.}\ }\textbf {\bibinfo
  {volume} {60}},\ \bibinfo {pages} {1351} (\bibinfo {year}
  {1988})}\BibitemShut {NoStop}%
\bibitem [{\citenamefont {Drummond}\ \emph {et~al.}(2014)\citenamefont
  {Drummond}, \citenamefont {Opanchuk}, \citenamefont {Rosales-Z\'{a}rate},\
  and\ \citenamefont {Reid}}]{Drummond2014-bell-sim}%
  \BibitemOpen
  \bibfield  {author} {\bibinfo {author} {\bibfnamefont {P.~D.}\ \bibnamefont
  {Drummond}}, \bibinfo {author} {\bibfnamefont {B.}~\bibnamefont {Opanchuk}},
  \bibinfo {author} {\bibfnamefont {L.~E.~C.}\ \bibnamefont
  {Rosales-Z\'{a}rate}}, \ and\ \bibinfo {author} {\bibfnamefont {M.~D.}\
  \bibnamefont {Reid}},\ }\href {\doibase 10.1088/0031-8949/2014/T160/014009}
  {\bibfield  {journal} {\bibinfo  {journal} {Phys. Scripta}\ }\textbf
  {\bibinfo {volume} {T160}},\ \bibinfo {pages} {014009} (\bibinfo {year}
  {2014})}\BibitemShut {NoStop}%
\bibitem [{\citenamefont {Opanchuk}\ \emph {et~al.}(2014)\citenamefont
  {Opanchuk}, \citenamefont {Rosales-Z\'{a}rate}, \citenamefont {Reid},\ and\
  \citenamefont {Drummond}}]{Opanchuk2014-bell-sim}%
  \BibitemOpen
  \bibfield  {author} {\bibinfo {author} {\bibfnamefont {B.}~\bibnamefont
  {Opanchuk}}, \bibinfo {author} {\bibfnamefont {L.~E.~C.}\ \bibnamefont
  {Rosales-Z\'{a}rate}}, \bibinfo {author} {\bibfnamefont {M.~D.}\ \bibnamefont
  {Reid}}, \ and\ \bibinfo {author} {\bibfnamefont {P.~D.}\ \bibnamefont
  {Drummond}},\ }\href {\doibase 10.1016/j.physleta.2014.01.045} {\bibfield
  {journal} {\bibinfo  {journal} {Phys. Lett. A}\ }\textbf {\bibinfo {volume}
  {378}},\ \bibinfo {pages} {946} (\bibinfo {year} {2014})}\BibitemShut
  {NoStop}%
\bibitem [{\citenamefont {Husimi}(1940)}]{Husimi1940}%
  \BibitemOpen
  \bibfield  {author} {\bibinfo {author} {\bibfnamefont {K.}~\bibnamefont
  {Husimi}},\ }\href
  {https://www.jstage.jst.go.jp/article/ppmsj1919/22/4/22\_4\_264/\_article}
  {\bibfield  {journal} {\bibinfo  {journal} {Proc. Phys. Math. Soc. Jpn.}\
  }\textbf {\bibinfo {volume} {22}},\ \bibinfo {pages} {264} (\bibinfo {year}
  {1940})}\BibitemShut {NoStop}%
\bibitem [{\citenamefont {Arecchi}\ \emph {et~al.}(1972)\citenamefont
  {Arecchi}, \citenamefont {Courtens}, \citenamefont {Gilmore},\ and\
  \citenamefont {Thomas}}]{Arecchi1972}%
  \BibitemOpen
  \bibfield  {author} {\bibinfo {author} {\bibfnamefont {F.}~\bibnamefont
  {Arecchi}}, \bibinfo {author} {\bibfnamefont {E.}~\bibnamefont {Courtens}},
  \bibinfo {author} {\bibfnamefont {R.}~\bibnamefont {Gilmore}}, \ and\
  \bibinfo {author} {\bibfnamefont {H.}~\bibnamefont {Thomas}},\ }\href
  {\doibase 10.1103/PhysRevA.6.2211} {\bibfield  {journal} {\bibinfo  {journal}
  {Phys. Rev. A}\ }\textbf {\bibinfo {volume} {6}},\ \bibinfo {pages} {2211}
  (\bibinfo {year} {1972})}\BibitemShut {NoStop}%
\bibitem [{\citenamefont {Gilmore}\ \emph {et~al.}(1975)\citenamefont
  {Gilmore}, \citenamefont {Bowden},\ and\ \citenamefont
  {Narducci}}]{Gilmore1975}%
  \BibitemOpen
  \bibfield  {author} {\bibinfo {author} {\bibfnamefont {R.}~\bibnamefont
  {Gilmore}}, \bibinfo {author} {\bibfnamefont {C.}~\bibnamefont {Bowden}}, \
  and\ \bibinfo {author} {\bibfnamefont {L.}~\bibnamefont {Narducci}},\ }\href
  {\doibase 10.1103/PhysRevA.12.1019} {\bibfield  {journal} {\bibinfo
  {journal} {Phys. Rev. A}\ }\textbf {\bibinfo {volume} {12}},\ \bibinfo
  {pages} {1019} (\bibinfo {year} {1975})}\BibitemShut {NoStop}%
\bibitem [{\citenamefont {Drummond}(1983)}]{Drummond1983}%
  \BibitemOpen
  \bibfield  {author} {\bibinfo {author} {\bibfnamefont {P.~D.}\ \bibnamefont
  {Drummond}},\ }\href {\doibase 10.1103/PhysRevLett.50.1407} {\bibfield
  {journal} {\bibinfo  {journal} {Phys. Rev. Lett.}\ }\textbf {\bibinfo
  {volume} {50}},\ \bibinfo {pages} {1407} (\bibinfo {year}
  {1983})}\BibitemShut {NoStop}%
\bibitem [{\citenamefont {Ardehali}(1992)}]{Ardehali1992}%
  \BibitemOpen
  \bibfield  {author} {\bibinfo {author} {\bibfnamefont {M.}~\bibnamefont
  {Ardehali}},\ }\href {\doibase 10.1103/PhysRevA.46.5375} {\bibfield
  {journal} {\bibinfo  {journal} {Phys. Rev. A}\ }\textbf {\bibinfo {volume}
  {46}},\ \bibinfo {pages} {5375} (\bibinfo {year} {1992})}\BibitemShut
  {NoStop}%
\bibitem [{\citenamefont {Belinski\u{\i}}\ and\ \citenamefont
  {Klyshko}(1993)}]{Belinskii1993-interference}%
  \BibitemOpen
  \bibfield  {author} {\bibinfo {author} {\bibfnamefont {A.~V.}\ \bibnamefont
  {Belinski\u{\i}}}\ and\ \bibinfo {author} {\bibfnamefont {D.~N.}\
  \bibnamefont {Klyshko}},\ }\href {\doibase 10.1070/PU1993v036n08ABEH002299}
  {\bibfield  {journal} {\bibinfo  {journal} {Phys-Usp.}\ }\textbf {\bibinfo
  {volume} {36}},\ \bibinfo {pages} {653} (\bibinfo {year} {1993})}\BibitemShut
  {NoStop}%
\bibitem [{\citenamefont {Belinsky}\ and\ \citenamefont
  {Klyshko}(1993)}]{Belinsky1993-N-particle}%
  \BibitemOpen
  \bibfield  {author} {\bibinfo {author} {\bibfnamefont {A.~V.}\ \bibnamefont
  {Belinsky}}\ and\ \bibinfo {author} {\bibfnamefont {D.~N.}\ \bibnamefont
  {Klyshko}},\ }\href {\doibase 10.1016/0375-9601(93)90471-B} {\bibfield
  {journal} {\bibinfo  {journal} {Phys. Lett. A}\ }\textbf {\bibinfo {volume}
  {176}},\ \bibinfo {pages} {415} (\bibinfo {year} {1993})}\BibitemShut
  {NoStop}%
\bibitem [{\citenamefont {Svetlichny}(1987)}]{Svetlichny1987}%
  \BibitemOpen
  \bibfield  {author} {\bibinfo {author} {\bibfnamefont {G.}~\bibnamefont
  {Svetlichny}},\ }\href {\doibase 10.1103/PhysRevD.35.3066} {\bibfield
  {journal} {\bibinfo  {journal} {Phys. Rev. D}\ }\textbf {\bibinfo {volume}
  {35}},\ \bibinfo {pages} {3066} (\bibinfo {year} {1987})}\BibitemShut
  {NoStop}%
\bibitem [{\citenamefont {Collins}\ \emph {et~al.}(2002)\citenamefont
  {Collins}, \citenamefont {Gisin}, \citenamefont {Popescu}, \citenamefont
  {Roberts},\ and\ \citenamefont {Scarani}}]{Collins2002}%
  \BibitemOpen
  \bibfield  {author} {\bibinfo {author} {\bibfnamefont {D.}~\bibnamefont
  {Collins}}, \bibinfo {author} {\bibfnamefont {N.}~\bibnamefont {Gisin}},
  \bibinfo {author} {\bibfnamefont {S.}~\bibnamefont {Popescu}}, \bibinfo
  {author} {\bibfnamefont {D.}~\bibnamefont {Roberts}}, \ and\ \bibinfo
  {author} {\bibfnamefont {V.}~\bibnamefont {Scarani}},\ }\href {\doibase
  10.1103/PhysRevLett.88.170405} {\bibfield  {journal} {\bibinfo  {journal}
  {Phys. Rev. Lett.}\ }\textbf {\bibinfo {volume} {88}},\ \bibinfo {pages}
  {170405} (\bibinfo {year} {2002})}\BibitemShut {NoStop}%
\bibitem [{\citenamefont {Ghose}\ \emph {et~al.}(2009)\citenamefont {Ghose},
  \citenamefont {Sinclair}, \citenamefont {Debnath}, \citenamefont {Rungta},\
  and\ \citenamefont {Stock}}]{Ghose2009}%
  \BibitemOpen
  \bibfield  {author} {\bibinfo {author} {\bibfnamefont {S.}~\bibnamefont
  {Ghose}}, \bibinfo {author} {\bibfnamefont {N.}~\bibnamefont {Sinclair}},
  \bibinfo {author} {\bibfnamefont {S.}~\bibnamefont {Debnath}}, \bibinfo
  {author} {\bibfnamefont {P.}~\bibnamefont {Rungta}}, \ and\ \bibinfo {author}
  {\bibfnamefont {R.}~\bibnamefont {Stock}},\ }\href {\doibase
  10.1103/PhysRevLett.102.250404} {\bibfield  {journal} {\bibinfo  {journal}
  {Phys. Rev. Lett.}\ }\textbf {\bibinfo {volume} {102}},\ \bibinfo {pages}
  {250404} (\bibinfo {year} {2009})}\BibitemShut {NoStop}%
\bibitem [{\citenamefont {Ajoy}\ and\ \citenamefont {Rungta}(2010)}]{Ajoy2010}%
  \BibitemOpen
  \bibfield  {author} {\bibinfo {author} {\bibfnamefont {A.}~\bibnamefont
  {Ajoy}}\ and\ \bibinfo {author} {\bibfnamefont {P.}~\bibnamefont {Rungta}},\
  }\href {\doibase 10.1103/PhysRevA.81.052334} {\bibfield  {journal} {\bibinfo
  {journal} {Phys. Rev. A}\ }\textbf {\bibinfo {volume} {81}},\ \bibinfo
  {pages} {052334} (\bibinfo {year} {2010})}\BibitemShut {NoStop}%
\bibitem [{\citenamefont {Bancal}\ \emph {et~al.}(2011)\citenamefont {Bancal},
  \citenamefont {Brunner}, \citenamefont {Gisin},\ and\ \citenamefont
  {Liang}}]{Bancal2011}%
  \BibitemOpen
  \bibfield  {author} {\bibinfo {author} {\bibfnamefont {J.-D.}\ \bibnamefont
  {Bancal}}, \bibinfo {author} {\bibfnamefont {N.}~\bibnamefont {Brunner}},
  \bibinfo {author} {\bibfnamefont {N.}~\bibnamefont {Gisin}}, \ and\ \bibinfo
  {author} {\bibfnamefont {Y.-C.}\ \bibnamefont {Liang}},\ }\href {\doibase
  10.1103/PhysRevLett.106.020405} {\bibfield  {journal} {\bibinfo  {journal}
  {Phys. Rev. Lett.}\ }\textbf {\bibinfo {volume} {106}},\ \bibinfo {pages}
  {020405} (\bibinfo {year} {2011})}\BibitemShut {NoStop}%
\bibitem [{\citenamefont {Chen}\ \emph {et~al.}(2011)\citenamefont {Chen},
  \citenamefont {Deng}, \citenamefont {Su}, \citenamefont {Wu},\ and\
  \citenamefont {Oh}}]{Chen2011}%
  \BibitemOpen
  \bibfield  {author} {\bibinfo {author} {\bibfnamefont {J.-L.}\ \bibnamefont
  {Chen}}, \bibinfo {author} {\bibfnamefont {D.-L.}\ \bibnamefont {Deng}},
  \bibinfo {author} {\bibfnamefont {H.-Y.}\ \bibnamefont {Su}}, \bibinfo
  {author} {\bibfnamefont {C.}~\bibnamefont {Wu}}, \ and\ \bibinfo {author}
  {\bibfnamefont {C.~H.}\ \bibnamefont {Oh}},\ }\href {\doibase
  10.1103/PhysRevA.83.022316} {\bibfield  {journal} {\bibinfo  {journal} {Phys.
  Rev. A}\ }\textbf {\bibinfo {volume} {83}},\ \bibinfo {pages} {022316}
  (\bibinfo {year} {2011})}\BibitemShut {NoStop}%
\bibitem [{\citenamefont {Grandjean}\ \emph {et~al.}(2012)\citenamefont
  {Grandjean}, \citenamefont {Liang}, \citenamefont {Bancal}, \citenamefont
  {Brunner},\ and\ \citenamefont {Gisin}}]{Grandjean2012}%
  \BibitemOpen
  \bibfield  {author} {\bibinfo {author} {\bibfnamefont {B.}~\bibnamefont
  {Grandjean}}, \bibinfo {author} {\bibfnamefont {Y.-C.}\ \bibnamefont
  {Liang}}, \bibinfo {author} {\bibfnamefont {J.-D.}\ \bibnamefont {Bancal}},
  \bibinfo {author} {\bibfnamefont {N.}~\bibnamefont {Brunner}}, \ and\
  \bibinfo {author} {\bibfnamefont {N.}~\bibnamefont {Gisin}},\ }\href
  {\doibase 10.1103/PhysRevA.85.052113} {\bibfield  {journal} {\bibinfo
  {journal} {Phys. Rev. A}\ }\textbf {\bibinfo {volume} {85}},\ \bibinfo
  {pages} {052113} (\bibinfo {year} {2012})}\BibitemShut {NoStop}%
\bibitem [{\citenamefont {Werner}\ and\ \citenamefont
  {Wolf}(2001)}]{Werner2001}%
  \BibitemOpen
  \bibfield  {author} {\bibinfo {author} {\bibfnamefont {R.~F.}\ \bibnamefont
  {Werner}}\ and\ \bibinfo {author} {\bibfnamefont {M.~M.}\ \bibnamefont
  {Wolf}},\ }\href {\doibase 10.1103/PhysRevA.64.032112} {\bibfield  {journal}
  {\bibinfo  {journal} {Phys. Rev. A}\ }\textbf {\bibinfo {volume} {64}},\
  \bibinfo {pages} {032112} (\bibinfo {year} {2001})}\BibitemShut {NoStop}%
\bibitem [{\citenamefont {Einstein}\ \emph {et~al.}(1935)\citenamefont
  {Einstein}, \citenamefont {Podolsky},\ and\ \citenamefont
  {Rosen}}]{Einstein1935}%
  \BibitemOpen
  \bibfield  {author} {\bibinfo {author} {\bibfnamefont {A.}~\bibnamefont
  {Einstein}}, \bibinfo {author} {\bibfnamefont {B.}~\bibnamefont {Podolsky}},
  \ and\ \bibinfo {author} {\bibfnamefont {N.}~\bibnamefont {Rosen}},\ }\href
  {\doibase 10.1103/PhysRev.47.777} {\bibfield  {journal} {\bibinfo  {journal}
  {Phys. Rev.}\ }\textbf {\bibinfo {volume} {47}},\ \bibinfo {pages} {777}
  (\bibinfo {year} {1935})}\BibitemShut {NoStop}%
\bibitem [{\citenamefont {Mermin}(1990{\natexlab{b}})}]{Mermin1990-reality}%
  \BibitemOpen
  \bibfield  {author} {\bibinfo {author} {\bibfnamefont {N.~D.}\ \bibnamefont
  {Mermin}},\ }\href {\doibase 10.1063/1.2810588} {\bibfield  {journal}
  {\bibinfo  {journal} {Phys. Today}\ }\textbf {\bibinfo {volume} {43}},\
  \bibinfo {pages} {9} (\bibinfo {year} {1990}{\natexlab{b}})}\BibitemShut
  {NoStop}%
\bibitem [{\citenamefont {Bell}(1964)}]{Bell1964}%
  \BibitemOpen
  \bibfield  {author} {\bibinfo {author} {\bibfnamefont {J.~S.}\ \bibnamefont
  {Bell}},\ }\href
  {http://philoscience.unibe.ch/documents/TexteHS10/bell1964epr.pdf} {\bibfield
   {journal} {\bibinfo  {journal} {Physics}\ }\textbf {\bibinfo {volume} {1}},\
  \bibinfo {pages} {195} (\bibinfo {year} {1964})}\BibitemShut {NoStop}%
\bibitem [{\citenamefont {Clauser}\ \emph {et~al.}(1969)\citenamefont
  {Clauser}, \citenamefont {Horne}, \citenamefont {Shimony},\ and\
  \citenamefont {Holt}}]{Clauser1969}%
  \BibitemOpen
  \bibfield  {author} {\bibinfo {author} {\bibfnamefont {J.~F.}\ \bibnamefont
  {Clauser}}, \bibinfo {author} {\bibfnamefont {M.~A.}\ \bibnamefont {Horne}},
  \bibinfo {author} {\bibfnamefont {A.}~\bibnamefont {Shimony}}, \ and\
  \bibinfo {author} {\bibfnamefont {R.~A.}\ \bibnamefont {Holt}},\ }\href
  {\doibase 10.1103/PhysRevLett.23.880} {\bibfield  {journal} {\bibinfo
  {journal} {Phys. Rev. Lett.}\ }\textbf {\bibinfo {volume} {23}},\ \bibinfo
  {pages} {880} (\bibinfo {year} {1969})}\BibitemShut {NoStop}%
\bibitem [{\citenamefont {Clauser}\ and\ \citenamefont
  {Shimony}(1978)}]{Clauser1978}%
  \BibitemOpen
  \bibfield  {author} {\bibinfo {author} {\bibfnamefont {J.~F.}\ \bibnamefont
  {Clauser}}\ and\ \bibinfo {author} {\bibfnamefont {A.}~\bibnamefont
  {Shimony}},\ }\href {\doibase 10.1088/0034-4885/41/12/002} {\bibfield
  {journal} {\bibinfo  {journal} {Rep. Prog. Phys.}\ }\textbf {\bibinfo
  {volume} {41}},\ \bibinfo {pages} {1881} (\bibinfo {year}
  {1978})}\BibitemShut {NoStop}%
\bibitem [{\citenamefont {D'Espagnat}(1971)}]{D'Espagnat1971}%
  \BibitemOpen
  \bibinfo {editor} {\bibfnamefont {B.}~\bibnamefont {D'Espagnat}},\ ed.,\
  \href@noop {} {\emph {\bibinfo {title} {{Foundations of Quantum
  Mechanics}}}}\ (\bibinfo  {publisher} {Academic Press},\ \bibinfo {address}
  {New York},\ \bibinfo {year} {1971})\BibitemShut {NoStop}%
\bibitem [{\citenamefont {Seevinck}\ and\ \citenamefont
  {Svetlichny}(2002)}]{Seevinck2002}%
  \BibitemOpen
  \bibfield  {author} {\bibinfo {author} {\bibfnamefont {M.}~\bibnamefont
  {Seevinck}}\ and\ \bibinfo {author} {\bibfnamefont {G.}~\bibnamefont
  {Svetlichny}},\ }\href {\doibase 10.1103/PhysRevLett.89.060401} {\bibfield
  {journal} {\bibinfo  {journal} {Phys. Rev. Lett.}\ }\textbf {\bibinfo
  {volume} {89}},\ \bibinfo {pages} {060401} (\bibinfo {year}
  {2002})}\BibitemShut {NoStop}%
\bibitem [{\citenamefont {Aolita}\ \emph {et~al.}(2012)\citenamefont {Aolita},
  \citenamefont {Gallego}, \citenamefont {Cabello},\ and\ \citenamefont
  {Ac\'{\i}n}}]{Aolita2012}%
  \BibitemOpen
  \bibfield  {author} {\bibinfo {author} {\bibfnamefont {L.}~\bibnamefont
  {Aolita}}, \bibinfo {author} {\bibfnamefont {R.}~\bibnamefont {Gallego}},
  \bibinfo {author} {\bibfnamefont {A.}~\bibnamefont {Cabello}}, \ and\
  \bibinfo {author} {\bibfnamefont {A.}~\bibnamefont {Ac\'{\i}n}},\ }\href
  {\doibase 10.1103/PhysRevLett.108.100401} {\bibfield  {journal} {\bibinfo
  {journal} {Phys. Rev. Lett.}\ }\textbf {\bibinfo {volume} {108}},\ \bibinfo
  {pages} {100401} (\bibinfo {year} {2012})}\BibitemShut {NoStop}%
\bibitem [{\citenamefont {Gallego}\ \emph {et~al.}(2012)\citenamefont
  {Gallego}, \citenamefont {W\"{u}rflinger}, \citenamefont {Ac\'{\i}n},\ and\
  \citenamefont {Navascu\'{e}s}}]{Gallego2012}%
  \BibitemOpen
  \bibfield  {author} {\bibinfo {author} {\bibfnamefont {R.}~\bibnamefont
  {Gallego}}, \bibinfo {author} {\bibfnamefont {L.~E.}\ \bibnamefont
  {W\"{u}rflinger}}, \bibinfo {author} {\bibfnamefont {A.}~\bibnamefont
  {Ac\'{\i}n}}, \ and\ \bibinfo {author} {\bibfnamefont {M.}~\bibnamefont
  {Navascu\'{e}s}},\ }\href {\doibase 10.1103/PhysRevLett.109.070401}
  {\bibfield  {journal} {\bibinfo  {journal} {Phys. Rev. Lett.}\ }\textbf
  {\bibinfo {volume} {109}},\ \bibinfo {pages} {070401} (\bibinfo {year}
  {2012})}\BibitemShut {NoStop}%
\bibitem [{\citenamefont {Bancal}\ \emph {et~al.}(2013)\citenamefont {Bancal},
  \citenamefont {Barrett}, \citenamefont {Gisin},\ and\ \citenamefont
  {Pironio}}]{Bancal2013}%
  \BibitemOpen
  \bibfield  {author} {\bibinfo {author} {\bibfnamefont {J.-D.}\ \bibnamefont
  {Bancal}}, \bibinfo {author} {\bibfnamefont {J.}~\bibnamefont {Barrett}},
  \bibinfo {author} {\bibfnamefont {N.}~\bibnamefont {Gisin}}, \ and\ \bibinfo
  {author} {\bibfnamefont {S.}~\bibnamefont {Pironio}},\ }\href {\doibase
  10.1103/PhysRevA.88.014102} {\bibfield  {journal} {\bibinfo  {journal} {Phys.
  Rev. A}\ }\textbf {\bibinfo {volume} {88}},\ \bibinfo {pages} {014102}
  (\bibinfo {year} {2013})}\BibitemShut {NoStop}%
\bibitem [{\citenamefont {Wigner}(1932)}]{Wigner1932}%
  \BibitemOpen
  \bibfield  {author} {\bibinfo {author} {\bibfnamefont {E.~P.}\ \bibnamefont
  {Wigner}},\ }\href {\doibase 10.1103/PhysRev.40.749} {\bibfield  {journal}
  {\bibinfo  {journal} {Phys. Rev.}\ }\textbf {\bibinfo {volume} {40}},\
  \bibinfo {pages} {749} (\bibinfo {year} {1932})}\BibitemShut {NoStop}%
\bibitem [{\citenamefont {Glauber}(1963)}]{Glauber1963-states}%
  \BibitemOpen
  \bibfield  {author} {\bibinfo {author} {\bibfnamefont {R.~J.}\ \bibnamefont
  {Glauber}},\ }\href {\doibase 10.1103/PhysRev.131.2766} {\bibfield  {journal}
  {\bibinfo  {journal} {Phys. Rev.}\ }\textbf {\bibinfo {volume} {131}},\
  \bibinfo {pages} {2766} (\bibinfo {year} {1963})}\BibitemShut {NoStop}%
\bibitem [{\citenamefont {Sudarshan}(1963)}]{Sudarshan1963}%
  \BibitemOpen
  \bibfield  {author} {\bibinfo {author} {\bibfnamefont {E.}~\bibnamefont
  {Sudarshan}},\ }\href {\doibase 10.1103/PhysRevLett.10.277} {\bibfield
  {journal} {\bibinfo  {journal} {Phys. Rev. Lett.}\ }\textbf {\bibinfo
  {volume} {10}},\ \bibinfo {pages} {277} (\bibinfo {year} {1963})}\BibitemShut
  {NoStop}%
\bibitem [{\citenamefont {Chaturvedi}\ \emph {et~al.}(1994)\citenamefont
  {Chaturvedi}, \citenamefont {Agarwal},\ and\ \citenamefont
  {Srinivasan}}]{Chaturvedi1994}%
  \BibitemOpen
  \bibfield  {author} {\bibinfo {author} {\bibfnamefont {S.}~\bibnamefont
  {Chaturvedi}}, \bibinfo {author} {\bibfnamefont {G.~S.}\ \bibnamefont
  {Agarwal}}, \ and\ \bibinfo {author} {\bibfnamefont {V.}~\bibnamefont
  {Srinivasan}},\ }\href {\doibase 10.1088/0305-4470/27/2/007} {\bibfield
  {journal} {\bibinfo  {journal} {J. Phys. A: Math. Gen.}\ }\textbf {\bibinfo
  {volume} {27}},\ \bibinfo {pages} {L39} (\bibinfo {year} {1994})}\BibitemShut
  {NoStop}%
\bibitem [{\citenamefont {de~Oliveira}(1992)}]{DeOliveira1992}%
  \BibitemOpen
  \bibfield  {author} {\bibinfo {author} {\bibfnamefont {F.~A.~M.}\
  \bibnamefont {de~Oliveira}},\ }\href {\doibase 10.1103/PhysRevA.45.5104}
  {\bibfield  {journal} {\bibinfo  {journal} {Phys. Rev. A}\ }\textbf {\bibinfo
  {volume} {45}},\ \bibinfo {pages} {5104} (\bibinfo {year}
  {1992})}\BibitemShut {NoStop}%
\bibitem [{\citenamefont {Corney}\ and\ \citenamefont
  {Drummond}(2003)}]{Corney2003}%
  \BibitemOpen
  \bibfield  {author} {\bibinfo {author} {\bibfnamefont {J.~F.}\ \bibnamefont
  {Corney}}\ and\ \bibinfo {author} {\bibfnamefont {P.~D.}\ \bibnamefont
  {Drummond}},\ }\href {\doibase 10.1103/PhysRevA.68.063822} {\bibfield
  {journal} {\bibinfo  {journal} {Phys. Rev. A}\ }\textbf {\bibinfo {volume}
  {68}},\ \bibinfo {pages} {063822} (\bibinfo {year} {2003})}\BibitemShut
  {NoStop}%
\bibitem [{\citenamefont {Radcliffe}(1971)}]{Radcliffe1971}%
  \BibitemOpen
  \bibfield  {author} {\bibinfo {author} {\bibfnamefont {J.~M.}\ \bibnamefont
  {Radcliffe}},\ }\href {\doibase 10.1088/0305-4470/4/3/009} {\bibfield
  {journal} {\bibinfo  {journal} {J. Phys. A: Gen. Phys.}\ }\textbf {\bibinfo
  {volume} {4}},\ \bibinfo {pages} {313} (\bibinfo {year} {1971})}\BibitemShut
  {NoStop}%
\bibitem [{\citenamefont {Zhang}\ \emph {et~al.}(1990)\citenamefont {Zhang},
  \citenamefont {Feng},\ and\ \citenamefont {Gilmore}}]{Zhang1990}%
  \BibitemOpen
  \bibfield  {author} {\bibinfo {author} {\bibfnamefont {W.-M.}\ \bibnamefont
  {Zhang}}, \bibinfo {author} {\bibfnamefont {D.~H.}\ \bibnamefont {Feng}}, \
  and\ \bibinfo {author} {\bibfnamefont {R.}~\bibnamefont {Gilmore}},\ }\href
  {\doibase 10.1103/RevModPhys.62.867} {\bibfield  {journal} {\bibinfo
  {journal} {Rev. Mod. Phys.}\ }\textbf {\bibinfo {volume} {62}},\ \bibinfo
  {pages} {867} (\bibinfo {year} {1990})}\BibitemShut {NoStop}%
\bibitem [{\citenamefont {Barry}\ and\ \citenamefont
  {Drummond}(2008)}]{Barry2008}%
  \BibitemOpen
  \bibfield  {author} {\bibinfo {author} {\bibfnamefont {D.~W.}\ \bibnamefont
  {Barry}}\ and\ \bibinfo {author} {\bibfnamefont {P.~D.}\ \bibnamefont
  {Drummond}},\ }\href {\doibase 10.1103/PhysRevA.78.052108} {\bibfield
  {journal} {\bibinfo  {journal} {Phys. Rev. A}\ }\textbf {\bibinfo {volume}
  {78}},\ \bibinfo {pages} {052108} (\bibinfo {year} {2008})}\BibitemShut
  {NoStop}%
\bibitem [{\citenamefont {Wootters}(1987)}]{Wootters1987}%
  \BibitemOpen
  \bibfield  {author} {\bibinfo {author} {\bibfnamefont {W.~K.}\ \bibnamefont
  {Wootters}},\ }\href {\doibase 10.1016/0003-4916(87)90176-X} {\bibfield
  {journal} {\bibinfo  {journal} {Ann. Phys.}\ }\textbf {\bibinfo {volume}
  {176}},\ \bibinfo {pages} {1} (\bibinfo {year} {1987})}\BibitemShut {NoStop}%
\bibitem [{\citenamefont {Wootters}\ and\ \citenamefont
  {Fields}(1989)}]{Wootters1989}%
  \BibitemOpen
  \bibfield  {author} {\bibinfo {author} {\bibfnamefont {W.~K.}\ \bibnamefont
  {Wootters}}\ and\ \bibinfo {author} {\bibfnamefont {B.~D.}\ \bibnamefont
  {Fields}},\ }\href {\doibase 10.1016/0003-4916(89)90322-9} {\bibfield
  {journal} {\bibinfo  {journal} {Ann. Phys.}\ }\textbf {\bibinfo {volume}
  {191}},\ \bibinfo {pages} {363} (\bibinfo {year} {1989})}\BibitemShut
  {NoStop}%
\bibitem [{\citenamefont {Gibbons}\ \emph {et~al.}(2004)\citenamefont
  {Gibbons}, \citenamefont {Hoffman},\ and\ \citenamefont
  {Wootters}}]{Gibbons2004}%
  \BibitemOpen
  \bibfield  {author} {\bibinfo {author} {\bibfnamefont {K.~S.}\ \bibnamefont
  {Gibbons}}, \bibinfo {author} {\bibfnamefont {M.~J.}\ \bibnamefont
  {Hoffman}}, \ and\ \bibinfo {author} {\bibfnamefont {W.~K.}\ \bibnamefont
  {Wootters}},\ }\href {\doibase 10.1103/PhysRevA.70.062101} {\bibfield
  {journal} {\bibinfo  {journal} {Phys. Rev. A}\ }\textbf {\bibinfo {volume}
  {70}},\ \bibinfo {pages} {062101} (\bibinfo {year} {2004})}\BibitemShut
  {NoStop}%
\bibitem [{\citenamefont {Wootters}(2004)}]{Wootters2004}%
  \BibitemOpen
  \bibfield  {author} {\bibinfo {author} {\bibfnamefont {W.~K.}\ \bibnamefont
  {Wootters}},\ }\href {\doibase 10.1147/rd.481.0099} {\bibfield  {journal}
  {\bibinfo  {journal} {IBM J. Res. Dev.}\ }\textbf {\bibinfo {volume} {48}},\
  \bibinfo {pages} {99} (\bibinfo {year} {2004})}\BibitemShut {NoStop}%
\bibitem [{\citenamefont {Wootters}(2006)}]{Wootters2006}%
  \BibitemOpen
  \bibfield  {author} {\bibinfo {author} {\bibfnamefont {W.~K.}\ \bibnamefont
  {Wootters}},\ }\href {\doibase 10.1007/s10701-005-9008-x} {\bibfield
  {journal} {\bibinfo  {journal} {Found. Phys.}\ }\textbf {\bibinfo {volume}
  {36}},\ \bibinfo {pages} {112} (\bibinfo {year} {2006})}\BibitemShut
  {NoStop}%
\bibitem [{\citenamefont {Bj\"{o}rk}\ \emph {et~al.}(2008)\citenamefont
  {Bj\"{o}rk}, \citenamefont {Klimov},\ and\ \citenamefont
  {S\'{a}nchez-Soto}}]{Bjork2008}%
  \BibitemOpen
  \bibfield  {author} {\bibinfo {author} {\bibfnamefont {G.}~\bibnamefont
  {Bj\"{o}rk}}, \bibinfo {author} {\bibfnamefont {A.~B.}\ \bibnamefont
  {Klimov}}, \ and\ \bibinfo {author} {\bibfnamefont {L.~L.}\ \bibnamefont
  {S\'{a}nchez-Soto}},\ }\href {\doibase 10.1016/S0079-6638(07)51007-3}
  {\bibfield  {journal} {\bibinfo  {journal} {Prog. Optics}\ }\textbf {\bibinfo
  {volume} {51}},\ \bibinfo {pages} {469} (\bibinfo {year} {2008})}\BibitemShut
  {NoStop}%
\bibitem [{\citenamefont {Biedenharn}\ and\ \citenamefont {van
  Dam}(1965)}]{Biedenharn1965}%
  \BibitemOpen
  \bibfield  {author} {\bibinfo {author} {\bibfnamefont {L.~C.}\ \bibnamefont
  {Biedenharn}}\ and\ \bibinfo {author} {\bibfnamefont {H.}~\bibnamefont {van
  Dam}},\ }\href@noop {} {\emph {\bibinfo {title} {{Quantum theory of angular
  momentum: a collection of reprints and original papers}}}}\ (\bibinfo
  {publisher} {Academic Press},\ \bibinfo {address} {New York},\ \bibinfo
  {year} {1965})\ p.\ \bibinfo {pages} {332}\BibitemShut {NoStop}%
\bibitem [{\citenamefont {Reid}\ and\ \citenamefont
  {Walls}(1986)}]{Reid1986-violations}%
  \BibitemOpen
  \bibfield  {author} {\bibinfo {author} {\bibfnamefont {M.~D.}\ \bibnamefont
  {Reid}}\ and\ \bibinfo {author} {\bibfnamefont {D.~F.}\ \bibnamefont
  {Walls}},\ }\href {\doibase 10.1103/PhysRevA.34.1260} {\bibfield  {journal}
  {\bibinfo  {journal} {Phys. Rev. A}\ }\textbf {\bibinfo {volume} {34}},\
  \bibinfo {pages} {1260} (\bibinfo {year} {1986})}\BibitemShut {NoStop}%
\bibitem [{\citenamefont {Brune}\ \emph {et~al.}(1996)\citenamefont {Brune}
  \emph {et~al.}}]{Brune1996}%
  \BibitemOpen
  \bibfield  {author} {\bibinfo {author} {\bibfnamefont {M.}~\bibnamefont
  {Brune}} \emph {et~al.},\ }\href {\doibase 10.1103/PhysRevLett.77.4887}
  {\bibfield  {journal} {\bibinfo  {journal} {Phys. Rev. Lett.}\ }\textbf
  {\bibinfo {volume} {77}},\ \bibinfo {pages} {4887} (\bibinfo {year}
  {1996})}\BibitemShut {NoStop}%
\bibitem [{\citenamefont {Monz}\ \emph {et~al.}(2011)\citenamefont {Monz} \emph
  {et~al.}}]{Monz2011}%
  \BibitemOpen
  \bibfield  {author} {\bibinfo {author} {\bibfnamefont {T.}~\bibnamefont
  {Monz}} \emph {et~al.},\ }\href {\doibase 10.1103/PhysRevLett.106.130506}
  {\bibfield  {journal} {\bibinfo  {journal} {Phys. Rev. Lett.}\ }\textbf
  {\bibinfo {volume} {106}},\ \bibinfo {pages} {130506} (\bibinfo {year}
  {2011})}\BibitemShut {NoStop}%
\end{thebibliography}%

\end{document}